\documentclass[journal]{IEEEtran}
\usepackage{amsmath,amsfonts,amsthm,amssymb,bm}
\usepackage{algorithmic}
\usepackage{array}
\usepackage[caption=false,font=normalsize,labelfont=sf,textfont=sf]{subfig}
\usepackage{textcomp}
\usepackage{stfloats}
\usepackage{url}
\usepackage{verbatim}
\usepackage{graphicx}
\usepackage{bbding}
\usepackage{multirow}
\usepackage{cite, bbm, xcolor}
\usepackage{refcount}
\usepackage{flafter}
\usepackage{makecell}
\usepackage{diagbox}
\usepackage{booktabs}
\usepackage{algorithm}

\newtheorem{remark}{Remark}
\theoremstyle{definition}

\usepackage{float}
\def\BibTeX{{\rm B\kern-.05em{\sc i\kern-.025em b}\kern-.08em
    T\kern-.1667em\lower.7ex\hbox{E}\kern-.125emX}}
    \expandafter\def\expandafter\normalsize\expandafter{%
    \normalsize%
    \setlength\abovedisplayskip{4pt}%
    \setlength\belowdisplayskip{4pt}%
    \setlength\abovedisplayshortskip{2pt}%
    \setlength\belowdisplayshortskip{2pt}%
}
\allowdisplaybreaks[4]
\usepackage{balance}

\begin{document}
\title{Environment Division Multiple Access Assisted Token Communications}
 \author{Deqiao Gan, Chongjun Ouyang, and Zhiguo Ding
         \thanks{D. Gan and Z. Ding are with the School of Electrical and Electronic Engineering, Nanyang Technological University, Singapore.}
         \thanks{C. Ouyang is with the School of Electronic Engineering and Computer Science, Queen Mary University of London, London E1 4NS, U.K.}
         }
\maketitle
\begin{abstract}
    Token communications enable compact source representations to be conveyed over wireless channels. However, tokens extracted solely from source content do not explicitly account for transmit-resource assignments or the resulting desired and interfering links. Environment division multiple access (EDMA) exploits spatially distributed antenna arrays and environment-dependent propagation characteristics to establish regions with reduced mutual interference. Each antenna array is treated as an independently controlled element, which can be assigned to a user stream based on its desired and interfering links. To exploit these characteristics, an EDMA-aware token communication framework is proposed, in which task-token encoding is coordinated with element assignments and digital beamforming. Specifically, action tokens are constructed to characterize the selected physical configurations and their resulting effective channels. These action tokens are then fused with task tokens through a residual network to generate EDMA-aware tokens, which are mapped into complex baseband symbol sequences for transmission. The task tokens are subsequently reconstructed at the receivers. Furthermore, two designs are developed to investigate the role of task information in physical configuration: i) a separate design, in which element assignments and beamformers are determined solely from environmental information; and ii) a joint design, in which task-token blocks are additionally exploited to refine these configurations. An average transmit-power minimization problem is formulated to jointly optimize the trainable token-processing parameters, element assignments, and digital beamformers subject to signal-to-interference-plus-noise ratio (SINR) and token-recovery distortion constraints. To address the resulting mixed-integer nonconvex problem, a two-stage procedure is developed. The token-processing functions are jointly trained offline, while element assignments and beamformers are updated online through a token-distortion-guided alternating optimization (TD-AO) algorithm. Numerical results demonstrate that the proposed framework reduces transmit power compared with conventional zero-forcing (ZF) and maximum-ratio transmission (MRT) schemes while maintaining stable token-recovery performance under varying blockage conditions. Moreover, additional power savings are achieved by the joint design over the separate design, particularly when multiple users compete for limited transmit resources.
\end{abstract}


\section{Introduction}

Transmitting high-dimensional source data over wireless links can incur substantial communication overhead, particularly when bandwidth and energy resources are limited. To reduce this burden, task-oriented communications learn compact representations that retain information needed for the intended task~\cite{Shao2022TaskOriented}.
Token communications provide a framework for exchanging compact representations as tokens and exploiting their contextual relationships during transmission and processing~\cite{Qiao2025TokenCommunications}. Their effectiveness depends on how well the information carried by the tokens is preserved during wireless transmission.
Therefore, efficient token transmission requires encoding and recovery mechanisms that account for channel impairments and available communication resources.

In~\cite{Bourtsoulatze2019DeepJSCC}, a neural encoder and decoder were jointly optimized through a differentiable channel model to minimize image-reconstruction distortion. This deep joint source--channel coding (JSCC) approach accounts for channel impairments during encoding and recovery. Propagating the reconstruction loss through the channel model allows the transmitted representation to be learned according to its recovery performance under channel noise. Beyond this channel-dependent design, the information content of the source representation can also guide the allocation of transmission resources.
The authors of~\cite{Dai2022NTSCC} used nonlinear transforms to obtain latent representations and trained an entropy model to characterize their statistical distribution. The estimated information content guided adaptive-rate transmission and linked resource allocation to reconstruction distortion.
For token-based sources, this principle requires decisions on which information to transmit and how to encode it. These decisions should exploit relationships among tokens and preserve reconstruction or task performance as channel conditions change.

To address these representation and transmission requirements, recent studies have investigated token-level encoding, adaptive token reduction, and context-assisted recovery~\cite{Ying2026TokenCommunications,Devoto2026AdaptiveToken,Erak2026AdaptiveTokenMerging,Shin2026ContextAwareToken,Zhang2026AdaTokenCom}. In~\cite{Ying2026TokenCommunications}, point tokens extracted from point-cloud data were mapped into digital constellation symbols through joint semantic-channel coding and modulation. A rate allocator and a channel adapter adjusted the transmitted representations according to source information and channel conditions. Beyond symbol-level encoding, token selection and compression provide additional means of controlling communication-resource use.
The authors of~\cite{Devoto2026AdaptiveToken} combined semantic token selection, embedding compression, and resource allocation to adjust both the token count and the embedding dimensions.
In~\cite{Erak2026AdaptiveTokenMerging}, similar tokens were merged without additional training. The merging thresholds were optimized to balance task accuracy, computation, and communication costs. Dependencies within token sequences can also be exploited at the receiver to reduce the information requiring transmission. In~\cite{Shin2026ContextAwareToken}, a shared masked language model supplied contextual priors for token detection, while the transmitter omitted highly predictable tokens through context-aware masking. A related use of token predictability was investigated in~\cite{Zhang2026AdaTokenCom}, which combined autoregressive prediction with arithmetic coding and receiver-side token generation. The source compression rate and modulation and coding scheme were jointly adapted to time-varying network conditions.
These studies have established mechanisms for adapting token representations and transmission rates to source information and channel conditions. In multiuser networks, the effectiveness of such adaptation also depends on the desired and interfering links associated with concurrent transmissions. Therefore, characterizing environment-dependent spatial relationships is important for identifying effective transmission conditions and coordinating user access.

To characterize these spatial relationships, environment division multiple access (EDMA) exploits spatially selective wireless propagation to support simultaneous multiuser transmission with reduced mutual interference~\cite{Ding2025EDMA,Ding2026AINativeEDMA}. In EDMA, spatially distributed antenna arrays are treated as \emph{independently controlled elements}, which can be assigned to different user streams according to their propagation characteristics. The desired spatial separation is achieved by exploiting differences in path loss, blockage, and interference leakage across the elements and users. Such an EDMA realization is not restricted to a particular antenna architecture. Conventional distributed antenna arrays can exploit existing propagation differences through element selection and beamforming, whereas flexible antenna technologies provide additional opportunities to configure the effective propagation environment. For instance, movable antennas and fluid antenna systems can adjust radiating positions to exploit favorable propagation conditions~\cite{Zhu2024MAModel,Wong2021FAS}, while pinching-antenna systems (PASS) enable configurable radiating positions along dielectric waveguides~\cite{Ding2026AINativeEDMA,Wang2025Modeling}. In particular, PASS-assisted EDMA can exploit location-dependent path losses and blockage to strengthen desired links and suppress inter-user interference~\cite{Ding2025EDMA}. Uplink and downlink transmission models based on such propagation characteristics were investigated in~\cite{Ding2025EDMA}, while the integration of flexible antennas and artificial intelligence (AI) into EDMA was discussed in~\cite{Ding2026AINativeEDMA}. Beyond the physical realization, effective EDMA configuration requires knowledge of the propagation environment and mechanisms for identifying favorable transmission conditions. To this end, channel knowledge maps associate channel information with transmitter and receiver locations to support environment-aware channel acquisition and communication design~\cite{Zeng2021CKM}. Site-specific learning further exploits environmental priors to infer suitable transmission configurations from available observations~\cite{Wang2026GenerativeSSBF,Zhao2026SiteSpecificMIMO}. In particular, conditional generative models were employed to generate beamforming candidates from coarse channel measurements in~\cite{Wang2026GenerativeSSBF}, while users' effective spatial subspaces were inferred for user grouping and reduced-dimensional channel acquisition in~\cite{Zhao2026SiteSpecificMIMO}. These approaches provide environmental knowledge and configuration-inference capabilities that can support the identification and exploitation of favorable propagation relationships in EDMA.

Since tokens are encoded into symbol sequences for wireless transmission, token-recovery quality depends on both the encoding strategy and the propagation conditions experienced by these sequences. Therefore, EDMA provides an opportunity to exploit environmental knowledge to guide the selection of desired and interfering links, while the resulting transmission conditions can guide the construction of symbol sequences from tokens. Three challenges arise in EDMA-assisted token communications. First, source-derived task tokens describe the content to be delivered, but their encoding must also account for the desired and interfering links induced by the EDMA configuration. Second, the EDMA configuration and token-recovery distortion are highly coupled. Changing the configuration alters the conditions under which the tokens are transmitted, while changing the token content or encoding can alter the recovery performance achieved by a given configuration.
Third, each user's EDMA configuration and transmitted tokens affect other users through interference. This coupling complicates the coordinated optimization of EDMA configurations under individual token-recovery requirements.

Motivated by the above challenges, an EDMA-aware token communication framework is proposed to coordinate token construction, token-to-symbol mapping and transmission, and token recovery with element assignments and digital beamforming. The main contributions are summarized as follows.
\begin{itemize}
    \item We propose an EDMA-aware framework for multiuser downlink token transmission. Specifically, we first construct action tokens that describe element assignments and effective desired and interfering channels induced by digital beamformers. A residual network fuses the action tokens with task tokens extracted from the source to generate EDMA-aware token blocks. These blocks are mapped into complex symbol sequences for transmission, and the receivers reconstruct the task-token blocks from the received signals over the frame.

    \item We develop separate and joint designs to examine how task information affects the selection of element assignments and digital beamformers. In the separate design, EDMA configurations are selected without observing the task-token realization. In the joint design, task-token blocks participate in selecting EDMA configurations. Both designs employ the same token mapping, transmission and recovery modules. Based on this framework, we formulate an average transmit-power minimization problem over the trainable token-processing parameters, element assignments, and digital beamformers, under signal-to-interference-plus-noise ratio (SINR) and token-distortion requirements.

    \item To address the resulting mixed-integer nonconvex problem, we develop a two-stage optimization framework that includes offline training and online optimization. At the offline stage, the action-token encoder, fusion functions, symbol mapper, and receivers are jointly trained to minimize task-token-recovery distortion over sampled task inputs and feasible EDMA configurations. At the online stage, with the trained parameters fixed, we develop a token-distortion-guided alternating optimization (TD-AO) algorithm for EDMA configuration refinement. The algorithm alternates between beamformer refinement and assignment updates. Each candidate configuration is evaluated through the complete token transmission and recovery process for the corresponding design.

    \item We provide numerical results to evaluate the proposed framework. The results show that both proposed designs require less average transmit power than conventional zero-forcing (ZF) and maximum-ratio transmission (MRT) schemes. The proposed framework also maintains lower token-recovery distortion than these schemes under varying blockage conditions. By matching EDMA configurations to the task-token blocks, the joint design achieves further power savings over the separate design, particularly when more users compete for limited transmit resources.
\end{itemize}

Section~\ref{sec:system_model} presents the system model. Section~\ref{sec:token_transmission} describes token construction, mapping, transmission, and recovery and formulates the optimization problem. Section~\ref{sec:algorithm} develops the offline training and online optimization procedures. Section~\ref{sec:numerical_results} presents numerical results, and Section~\ref{sec:conclusion} concludes the paper.

\subsubsection*{Notations}
Scalars, vectors, and matrices are denoted by plain, bold lowercase, and bold uppercase letters, respectively. The operators $(\cdot)^{\mathsf T}$ and $(\cdot)^{\mathsf H}$ denote transpose and Hermitian transpose. The symbols $|a|$, $\|\mathbf a\|_2$, and $\|\mathbf A\|_{\rm F}$ denote the magnitude of a scalar, the Euclidean norm of a vector, and the Frobenius norm of a matrix, respectively. The term $j\in\mathcal K_{\rm c}\setminus\{k\}$ means that $j \in \mathcal K_{\rm c}$ but $j \neq k$. The operator $\operatorname{vec}(\mathbf A)$ stacks the columns of $\mathbf A$, and $\mathbb E[\cdot]$ denotes expectation. For a positive integer $J$, $[J]\triangleq\{1,\ldots,J\}$. The notation $\mathcal{CN}(\boldsymbol\mu,\boldsymbol\Sigma)$ denotes a circularly symmetric complex Gaussian distribution with mean $\boldsymbol\mu$ and covariance $\boldsymbol\Sigma$. Finally, $[x]_+\triangleq\max\{x,0\}$, and $\mathcal O(\cdot)$ denotes asymptotic complexity.


\section{System Model}
\label{sec:system_model}
\begin{figure}[t]
    \centering
    \includegraphics[width=\columnwidth]{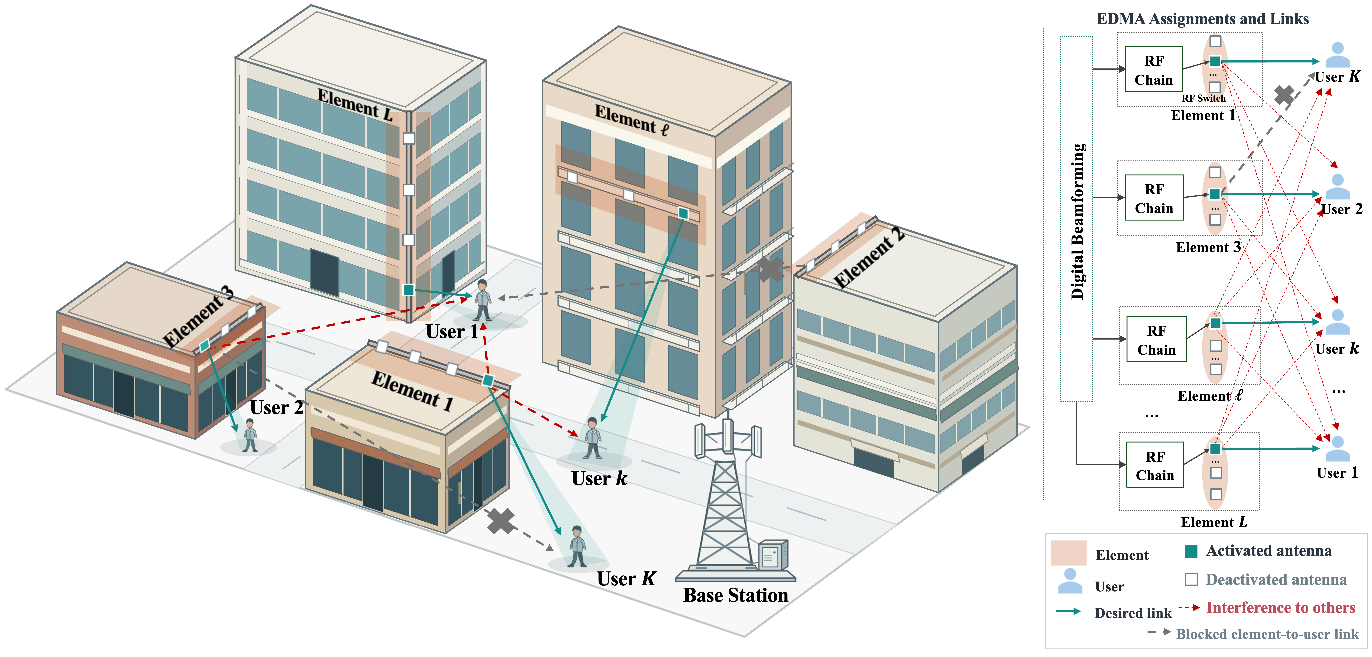}
    \caption{Proposed EDMA-aware token communication system with PASS as an example of a flexible-antenna implementation.}
    \label{system_model}
\end{figure}

Consider a downlink EDMA-enabled multiple-input single-output (MISO) system, where a central transmitter serves $K$ single-antenna users through $L$ independently driven transmit elements, as illustrated in Fig.~\ref{system_model}. One data stream is intended for each user. Each element $\ell\in[L]$ can be assigned to at most one user stream within a transmission interval.
The elements assigned to the same user stream form its environment division in the considered EDMA realization~\cite{Ding2025EDMA,Ding2026AINativeEDMA}. Each element is equipped with one radio-frequency (RF) chain and employs a single active radiating antenna, whose position can be configured through antenna selection or flexible-antenna techniques~\cite{Zhu2024MAModel,Wong2021FAS,Ouyang2026RateRegion}.

\subsection{Channel Model}
\label{subsec:channel_model}
Each frame is divided into $B_{\rm f}$ transmission intervals indexed by $\beta \in [B_{\rm f}]$. The channels are assumed to remain coherent over the frame. The prescribed antenna configurations $\rho_\beta$, user positions, and blockage states are fixed within each frame and may vary between frames. The configuration $\rho_\beta$ specifies the radiating position of each element.
The positions of user $k$ and the radiating antenna of element $\ell$ are denoted by $\mathbf u_{k,\beta}\in\mathbb R^3$ and $\mathbf u^{\rm e}_{\ell,\beta}(\rho_\beta) \in\mathbb R^3$, respectively. 
The position of the active radiating antenna serves as the spatial reference point for the element-to-user propagation channel.

We adopt a narrowband line-of-sight (LoS) channel model and neglect non-LoS (NLoS) components. Blockage completely suppresses the corresponding element-to-user link, and the channel coefficient from element $\ell$ to user $k$ is expressed as follows:
\begin{equation}
    \label{eq:element_user_channel}
    h_{k,\ell,\beta}(\rho_\beta,b_{k,\ell,\beta}) = b_{k,\ell,\beta}(\rho_\beta)\frac{\lambda e^{\mathrm i\frac{2\pi}{\lambda}\left\|\mathbf u_{k,\beta}-\mathbf u^{\rm e}_{\ell,\beta}(\rho_\beta)\right\|_2}}{4\pi \|\mathbf u_{k,\beta}-\mathbf u^{\rm e}_{\ell,\beta}(\rho_\beta)\|_2},
\end{equation}
where $\lambda$ is the carrier wavelength, the amplitude term assumes isotropic free-space propagation, and $b_{k,\ell,\beta}(\rho_\beta)\in\{0,1\}$ indicates direct-path visibility between the radiating antenna of element $\ell$ and user $k$. When the link is blocked, $b_{k,\ell,\beta}(\rho_\beta)=0$ sets $h_{k,\ell,\beta}(\rho_\beta, b_{k,\ell,\beta})=0$. If the direct propagation path is unobstructed, $b_{k,\ell,\beta}(\rho_\beta)=1$. Let $\mathbf b_{k,\beta}\triangleq[b_{k,1,\beta},\ldots,b_{k,L,\beta}]^{\mathsf T}\in\{0,1\}^{L}$ collect visibility indicators for user $k$, and let $\mathbf B_\beta\triangleq[\mathbf b_{1,\beta},\ldots,\mathbf b_{K,\beta}]^{\mathsf T}\in\{0,1\}^{K\times L}$ collect them for all users in interval $\beta$.
The channel coefficient varies nonlinearly with element position $\mathbf u^{\rm e}_{\ell,\beta}(\rho_\beta)$ through the distance-dependent amplitude and phase in~\eqref{eq:element_user_channel}. We assume the transmitter has perfect channel state information (CSI) $\mathbf h_{k,\beta}(\rho_\beta,\mathbf b_{k,\beta})$ for all $k\in[K]$ at the beginning of each frame.
The channel coefficient vector from all elements to user $k$ is collected as follows:
\begin{equation}
    \label{eq:user_channel_vector}
    \mathbf h_{k,\beta}(\rho_\beta,\mathbf b_{k,\beta}) \triangleq [h_{k,\ell,\beta}(\rho_\beta,b_{k,\ell,\beta})]_{\ell=1}^{L}\in{\mathbb{C}}^{L\times1}.
\end{equation}

\subsection{EDMA-Aware Action}
\label{subsec:edma_aware_action}

Let $\alpha_{k, \ell, \beta}\in\{0,1\}$ indicate whether element $\ell$ is assigned to the stream intended for user $k$ in interval $\beta$. Then, the element assignment vector for user $k$ can be expressed as $\mathbf a_{k,\beta}\triangleq [\alpha_{k,1,\beta},\ldots,\alpha_{k,L,\beta}]^{\mathsf T}\in\{0,1\}^{L}$. Assume $L \ge K$. In each interval, every element may be assigned to at most one user stream. The sum constraint $\sum_{j=1}^{K}\alpha_{j,\ell,\beta}\leq 1$, $\forall \ell\in[L]$ also permits an idle element. Consequently, the per-interval network-wide EDMA action representing assignments for all users can be expressed as follows:
\begin{equation}
    \label{eq:edma_action}
    \mathbf A_\beta \triangleq [\mathbf a_{1,\beta},\ldots,\mathbf a_{K,\beta}]\in\{0,1\}^{L\times K}.
\end{equation}

To assess the suitability of each element for the specific user, we define the channel power gain between element $\ell$ and user $k$ as follows:
\begin{equation}
    \label{eq:port_user_coupling}
    \kappa_{k,\ell,\beta}(\rho_\beta,b_{k,\ell,\beta}) \triangleq | h_{k,\ell,\beta}(\rho_\beta,b_{k,\ell,\beta})|^2.
\end{equation}

To assess the suitability of element $\ell$ for stream $j$, we also account for leakage to other users. The environment selectivity score $\eta_{j, \ell, \beta}$ is defined as the ratio of the desired-link power gain to the regularized aggregate leakage to the other users, which is expressed as follows:
\begin{equation}
    \label{eq:environment_selectivity}
    \eta_{j, \ell, \beta}(\rho_\beta, \mathbf B_{\beta}) \triangleq \frac{\kappa_{j,\ell,\beta}(\rho_\beta,b_{j,\ell,\beta})}{\sum_{k\ne j}\kappa_{k,\ell,\beta}(\rho_\beta,b_{k,\ell,\beta})+\epsilon_{\mathrm{reg}}},
\end{equation}
where the regularization constant $\epsilon_{\mathrm{reg}}>0$ prevents division by zero when the unintended links are blocked~\cite{HajHmida2021Pilot}. When $\eta_{j,\ell, \beta}>\eta_0$, element $\ell$ is suitable to serve user $j$, where $\eta_0$ is the selectivity threshold motivated by the signal-to-leakage-and-noise criterion in multiuser precoding~\cite{Sadek2007Leakage,Sadek2007Antenna}. A larger $\eta_{j, \ell, \beta}$ indicates a stronger desired link relative to interference leakage. This score guides the generation of feasible element assignments.

\subsection{Signal Model}
\label{subsec:signal_model}
For each user, task tokens are first extracted from the source content. The transmitter maps tokens intended for each user into complex baseband symbol sequences according to the selected EDMA assignment and digital beamformers. These sequences are spatially weighted by the beamformers and transmitted through assigned elements. Each user reconstructs its intended tokens from the received signal blocks. The token-to-symbol mapping and token recovery are detailed in Section~\ref{sec:token_transmission}.
Let $\mathbf v_{j,\beta} =[v_{j,1,\beta},\ldots,v_{j,L,\beta}]^{\mathsf T} \in\mathbb C^L$ denote the digital beamforming vector for stream $j$. Notably, $v_{j,\ell,\beta}$ must vanish if the element $\ell$ is not assigned to that stream, i.e., $v_{j,\ell,\beta}=0$ as long as $\alpha_{j, \ell, \beta}=0$. Thus, $\alpha_{j, \ell, \beta}$ specifies whether element $\ell$ may carry stream $j$, whereas $v_{j, \ell, \beta} \neq 0$ specifies the complex transmit weight applied at that element.
Let $\mathbf s_{j,\beta}\in\mathbb C^{1\times T_\beta}$ denote the baseband symbol sequence for user $j$ in interval $\beta$, which contains $T_\beta$ data channel uses.
Then, the transmitted signal block is expressed as follows:
\begin{equation}
    \label{eq:transmitted_block}
    \mathbf X_\beta = \sum_{j=1}^{K}\mathbf v_{j,\beta}\mathbf s_{j,\beta},
\end{equation}
where the $\ell$-th row of $\mathbf X_\beta$ contains symbols transmitted by element $\ell$ and $T_\beta$ is the number of data channel uses.

The received signal block at user $k$ is expressed as follows:
\begin{equation}
    \label{eq:received_block}
    \begin{aligned}
        \mathbf y_{k,\beta}= & g_{k\leftarrow k,\beta}(\rho_\beta,\mathbf b_{k,\beta})\mathbf s_{k,\beta} \\
        &+ \sum_{\!j\in [K]\setminus k\!} \!g_{k\leftarrow j,\beta}(\rho_\beta,\!\mathbf b_{k,\beta})\mathbf s_{j,\beta} + \mathbf n_{k,\beta},
    \end{aligned}
\end{equation}
where $\mathbf n_{k,\beta}\sim \mathcal{CN}(\mathbf{0},\sigma_k^2\mathbf{I}_{T_\beta})$ is additive white Gaussian noise (AWGN) with the variance $\sigma_k^2$. The effective coefficient from stream $j$ to user $k$ is expressed as follows:
\begin{equation}
    \label{eq:effective_coefficient}
    g_{k\leftarrow j,\beta}(\rho_\beta,\mathbf b_{k,\beta}) \triangleq \mathbf h_{k,\beta}^{\mathsf H}(\rho_\beta,\mathbf b_{k,\beta})\mathbf v_{j,\beta}.
\end{equation}

Moreover, $g_{k\leftarrow k,\beta}(\rho_\beta,\mathbf b_{k,\beta})\mathbf s_{k,\beta}$ is the desired signal for user $k$, while each term indexed by $j\ne k$ is the interference from another user's intended stream. Further, the received signal-to-interference-plus-noise ratio (SINR) of user $k$ in interval $\beta$ under unit-power reference streams is defined as follows:
\begin{equation}
    \label{eq:token_sinr}
    \gamma_{k,\beta}(\rho_\beta,\mathbf b_{k,\beta}) \triangleq \frac{|g_{k\leftarrow k,\beta}(\rho_\beta,\mathbf b_{k,\beta})|^{2}} {\sum_{j\in[K] \setminus k}|g_{k\leftarrow j,\beta}(\rho_\beta,\mathbf b_{k,\beta})|^{2} +\sigma_k^{2}}.
\end{equation}

\section{Joint and Separate Designs for EDMA-Aware Token Transmission}
\label{sec:token_transmission}



\begin{figure}[t]
    \centering
    \includegraphics[width=0.9\columnwidth]{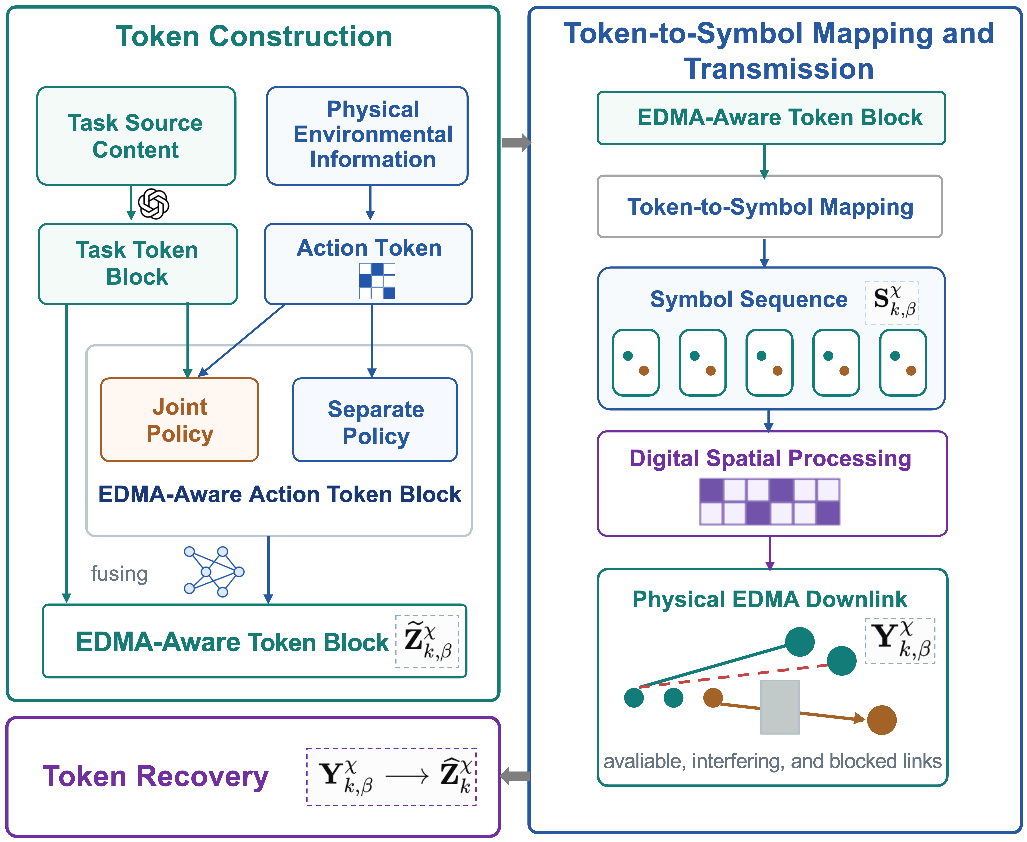}
    \caption{Illustration of the proposed EDMA-aware token communication framework for the separate and joint designs.}
    \label{framework}
\end{figure}

This section develops token construction, symbol mapping, and recovery under separate and joint designs. The two designs differ in the information used to select the EDMA assignment and digital beamformers.
The separate design selects the configurations from the available physical information alone. The joint design also uses the current task-token blocks.
Both designs follow the same token-construction, token-to-symbol mapping, and token recovery steps, as illustrated in Fig.~\ref{framework}. We then formulate the average transmit-power minimization problem for each design.

\subsection{EDMA-Aware Token Construction}
\label{subsec:token_construction}


The source content is encoded into compact vectors, termed \emph{task tokens}, to reduce transmission overhead. For each user, an \emph{action token} describes the selected element assignment and the effective desired and interfering channels. A residual network fuses the action token with each task token to generate EDMA-aware tokens for the token-to-symbol mapper.

\subsubsection{Task Token}
Let $\boldsymbol{\psi}_k$ denote the source content available at the transmitter for the task intended for user $k$, such as image pixels. The task encoder $\Phi_{\rm task}(\cdot)$ maps $\boldsymbol{\psi}_k$ to the task-token block $\mathbf Z_k=\Phi_{\rm task}(\boldsymbol{\psi}_k) \in\mathbb R^{N_k\times d_k}$. This block consists of $N_k$ task tokens and is expressed as follows:
\begin{equation}
    \label{eq:task_token_block}
    \mathbf Z_k = [\mathbf z_{k,1},\ldots,\mathbf z_{k,N_k}]^{\mathsf T}\in\mathbb R^{N_k\times d_k}, n \in [N_k],
\end{equation}
where the dimension of each task token is $d_k$.
The task token blocks for all users are denoted by $\mathcal Z\triangleq(\mathbf Z_1,\ldots,\mathbf Z_K)$.

\subsubsection{Action Token}
Equations~\eqref{eq:port_user_coupling} and~\eqref{eq:environment_selectivity} guide the selection of the element-assignment vector $\mathbf{a}_{k,\beta}$ for user $k$. The action encoder $\Phi_{\mathrm{act}}(\cdot)$ takes two inputs for user $k$: its element-assignment vector $\mathbf a_{k,\beta}$ and the effective channel coefficients $\{g_{k\leftarrow j,\beta}(\rho_\beta,\mathbf b_{k,\beta})\}_{j=1}^{K}$ from all streams to that user. Thus, the action token is expressed as follows:
\begin{equation}
    \label{eq:action_token}
    \mathbf q_{k,\beta} = \Phi_{\mathrm{act}}(\mathbf a_{k,\beta},\{g_{k\leftarrow j,\beta}(\rho_\beta,\mathbf b_{k,\beta})\}_{j=1}^{K}) \in\mathbb R^{d_q},
\end{equation}
where the same construction applies to the separate and joint designs using their corresponding assignments and beamformers.

\subsubsection{EDMA-Aware Token}

To determine whether the task content influences these physical decisions, we consider separate and joint designs, indexed by $\chi\in\{\mathrm{sep},\mathrm{joint}\}$. For each design $\chi$, the action token $\mathbf q_{k,\beta}^{\chi}$ is obtained from~\eqref{eq:action_token}.
A residual network fuses the task and action tokens through an additive correction and preserves a direct path for the task token. Given $\mathbf z_{k,n}$ and $\mathbf q^{\chi}_{k,\beta}$, the EDMA-aware token is expressed as follows:
\begin{equation}
    \label{eq:conditional_token_fusion}
    \widetilde{\mathbf z}^{\chi}_{k,n,\beta} \!= \!\mathbf z_{k,n}\! + \!\Phi_{\rm residual}(\mathbf z_{k,n},\Phi_{\rm EDMA}(\mathbf q^{\chi}_{k,\beta})),
\end{equation}
where $\Phi_{\rm EDMA}(\cdot)$ maps the action token from $\mathbb R^{d_q}$ to $\mathbb R^{d_k}$, and $\Phi_{\rm residual}(\cdot)$ computes the additive correction. The task token $\mathbf z_{k,n}$ provides the direct path in the residual structure.
\begin{remark}
The residual structure preserves a direct path for the task token and applies an action-dependent correction. Alternative fusion methods include concatenation followed by a multilayer perceptron (MLP).
\end{remark}
Thus, the EDMA-aware token blocks under the two designs are stacked as follows:
\begin{equation}
    \label{eq:edma_aware_token_block}
    \widetilde{\mathbf Z}^{\chi}_{k,\beta}\triangleq[\widetilde{\mathbf z}^{\chi}_{k,1,\beta},\ldots,\widetilde{\mathbf z}^{\chi}_{k,N_k,\beta}]^{\mathsf T}\in\mathbb R^{N_k\times d_k}.
\end{equation}

Let $\mathcal I_{\rm tx}\triangleq (\{\mathbf h_{k,\beta}^{\mathsf H}\}_{k\in[K]},\rho_\beta, \mathbf{B}_\beta, \mathbf u^{\rm e}_{\ell,\beta}, \mathbf u_{k,\beta})$ collect the transmitter's environmental information over the frame, including element-to-user channels, antenna configurations, blockage, and user and element radiating positions for all intervals. Let $\mathbf V_\beta^\chi\triangleq[\mathbf v_{1,\beta}^\chi,\ldots,\mathbf v_{K,\beta}^\chi]\in\mathbb C^{L\times K}$ collect the digital beamformers.
Both designs use the same feasible assignment and beamformer sets. They differ in whether the current task information is available during action selection.

\emph{i) Separate Design:}
The separate design follows the following selection policy:
\begin{equation}
    \label{eq:separate_policy}
    \pi_\beta^{\rm sep}:\mathcal I_{\rm tx} \longmapsto (\mathbf A_\beta^{\rm sep},\mathbf V_\beta^{\rm sep}),
\end{equation}
specifically, $\pi_\beta^{\rm sep}$ selects EDMA assignment $\mathbf A_\beta^{\rm sep}$ and digital beamformers $\mathbf V_\beta^{\rm sep}$ using environmental information $\mathcal I_{\rm tx}$, without using task-token blocks $\mathcal Z$. The effective coefficients $\{g_{k\leftarrow j,\beta}^{\rm sep} (\rho_\beta,\mathbf b_{k,\beta})\}_{j=1}^{K}=\{g_{k\leftarrow j,\beta}(\rho_\beta,\mathbf b_{k,\beta})\}_{j=1}^{K}$ are then computed from~\eqref{eq:effective_coefficient}. The corresponding action token is formulated as $\mathbf q_{k,\beta}^{\rm sep}= \Phi_{\rm act}(\mathbf a_{k,\beta}^{\rm sep},\{g_{k\leftarrow j,\beta}^{\rm sep}(\rho_\beta,\mathbf b_{k,\beta})\}_{j=1}^{K})$, see~\eqref{eq:action_token}.
Equation~\eqref{eq:conditional_token_fusion} fuses this action token with each task token $\mathbf z_{k,n}$ to obtain $\widetilde{\mathbf z}_{k,n,\beta}^{\rm sep}$. The resulting tokens are stacked into $\widetilde{\mathbf Z}_{k,\beta}^{\rm sep}$ according to~\eqref{eq:edma_aware_token_block}. Thus, task-token fusion follows action-token construction.

\emph{ii) Joint Design:}
The joint design follows the following selection policy:
\begin{equation}
    \label{eq:joint_policy}
    \pi_\beta^{\rm joint}:(\mathcal I_{\rm tx},\mathcal Z)\longmapsto(\mathbf A_\beta^{\rm joint},\mathbf V_{\beta}^{\rm joint}),
\end{equation}
specifically, $\pi_\beta^{\rm joint}$ uses both environmental information $\mathcal I_{\rm tx}$ and task-token blocks $\mathcal Z$ to select EDMA assignment $\mathbf A_\beta^{\rm joint}$ and digital beamformers $\mathbf V_\beta^{\rm joint}$. The physical configurations are first selected from $\mathcal I_{\rm tx}$, as in the separate design. The joint policy then refines these configurations using the current task-token blocks $\mathcal Z$. The corresponding action token for user $k$ is given by
\begin{equation}
    \label{eq:joint_action_token}
    \mathbf q_{k,\beta}^{\rm joint}=\Phi_{\rm act}(\mathbf a_{k,\beta}^{\rm joint},\{g_{k\leftarrow j,\beta}^{\rm joint}(\rho_\beta,\mathbf b_{k,\beta})\}_{j=1}^{K}),
\end{equation}
where the effective coefficients are determined by the joint selection policy through~\eqref{eq:effective_coefficient}.
Substituting $\mathbf q_{k,\beta}^{\rm joint}$ into~\eqref{eq:conditional_token_fusion} gives $\widetilde{\mathbf z}_{k,n,\beta}^{\rm joint}$, and the resulting block $\widetilde{\mathbf Z}_{k,\beta}^{\rm joint}$ is formed using~\eqref{eq:edma_aware_token_block}. Thus, task tokens participate in both physical-configuration selection and subsequent fusion under the joint design.

\subsection{Token-to-Symbol Mapping and Transmission}
\label{subsec:mapping}

For either design $\chi$, EDMA-aware token blocks are mapped into a complex baseband symbol sequence as follows:
\begin{equation}
    \label{eq:token_symbol_mapping}
    \mathbf s_{k,\beta}^{\chi}=F_{\mu}(\widetilde{\mathbf Z}_{k,\beta}^{\chi})\in\mathbb C^{1\times T_\beta},
\end{equation}
where $F_\mu(\cdot)$ is a token-to-symbol mapping function with trainable parameter $\mu$.
The mapper implements token-level JSCC through a direct mapping from token features to channel input symbols. Its parameters are trained according to the deep JSCC principle~\cite{Bourtsoulatze2019DeepJSCC}.
Specifically, the learnable transformation produces $2T_\beta$ real-valued outputs. The first $T_\beta$ outputs specify the real components of the symbol sequence, and the remaining $T_\beta$ outputs specify its imaginary components. The resulting sequence contains $T_\beta$ complex symbols and occupies $T_\beta$ channel uses. Each output sequence is normalized to unit average symbol energy, i.e., $\|\mathbf s_{k,\beta}^{\chi}\|_2^2=T_\beta$. The real and imaginary components of symbols are learned directly, without imposing a predefined finite constellation.
For either design $\chi$, the resulting symbol sequences are transmitted using selected digital beamformers. The transmitted and received signal blocks follow \eqref{eq:transmitted_block} and \eqref{eq:received_block}, respectively.

\subsection{Token Recovery}

For either design $\chi\in\{\mathrm{sep},\mathrm{joint}\}$, user $k$ reconstructs its task-token block from the received signals. We assume perfect effective CSI at each receiver. The receiver-side information comprises the effective channel coefficients and the noise variance as $\mathcal I_{{\rm rx},k}^{\chi}\triangleq (\{g_{k\leftarrow j,\beta}^{\chi} (\rho_\beta,\mathbf b_{k,\beta})\}_{j\in[K],\,\beta\in[B_{\rm f}]}, \sigma_k^2)$.
Utilizing $\mathcal I_{{\rm rx},k}^{\chi}$, user $k$ decodes the received blocks to recover $\widehat{\mathbf Z}^{\chi}_k$, which is given by
\begin{equation}
    \label{eq:task_token_recovery}
    \widehat{\mathbf Z}_{k}^{\chi}=D_{\phi,k}(\{\mathbf y_{k,\beta}^{\chi}\}_{\beta=1}^{B_{\rm f}},\mathcal I_{\mathrm{rx},k}^{\chi}),
\end{equation}
where $D_{\phi,k}(\cdot)$ denotes the receiver decoder, and $\phi$ collects its trainable parameters. The transmitter-side action token $\mathbf q_{k,\beta}^{\chi}$ determines the transmitted symbol sequences. The receiver reconstructs the task-token block for the intended service; separate action-token recovery is not required. Recovery performance is measured by the distortion between $\mathbf Z_k$ and $\widehat{\mathbf Z}_k^{\chi}$, which is defined as follows:
\begin{equation}
    \label{eq:task_token_distortion}
    \Delta_k^{\chi} \triangleq \frac{1}{N_k d_k} \mathbb E [\|\mathbf Z_k-\widehat{\mathbf Z}_k^{\chi}\|_{\mathrm F}^{2}],
\end{equation}
where normalization by $N_kd_k$ gives the mean squared error averaged over all task-token components. The expectation is taken over task-source samples, propagation, and receiver noise, with assignments and beamformers determined by the corresponding selection policy.

\subsection{Problem Formulation}
\label{subsec:problem_formulation}


The selected element assignments and digital beamformers determine the effective desired and interfering links. They also affect symbol mapping through the action tokens and thus influence task-token recovery. We therefore minimize transmit power subject to each user's SINR and token-distortion constraints.

Let $T_{\rm f} \triangleq \sum_{\beta=1}^{B_{\rm f}}T_\beta$ denote the total number of data channel uses per frame. The average transmit power is defined as follows:
\begin{equation}
    \label{eq:average_transmit_power}
    \bar P^\chi \triangleq \frac{1}{T_{\rm f}} \sum_{\beta=1}^{B_{\rm f}}\mathbb E[\|\mathbf X_\beta^\chi\|_{\rm F}^{2}],
\end{equation}
where the exclusive beamformer supports and symbol normalization yield $\|\mathbf X_\beta^\chi\|_{\rm F}^{2} = T_\beta\sum_{j=1}^{K} \|\mathbf v_{j,\beta}^\chi\|_2^2$ based on $\|\mathbf s_{j,\beta}^{\chi}\|_2^2=T_\beta$.

Let $\boldsymbol{\vartheta}$ collect the trainable parameters of the action-token encoder $\Phi_{\rm act}(\cdot)$, the fusion functions $\Phi_{\rm EDMA}(\cdot)$ and $\Phi_{\rm residual}(\cdot)$, the symbol mapper $F_\mu(\cdot)$ with $\mu$, and receivers $\{D_{\phi,k}(\cdot)\}_{k=1}^{K}$ with $\phi$.
Therefore, the power-minimization problem for $\chi\in\{\mathrm{sep},\mathrm{joint}\}$ can be formulated as follows:
\begin{subequations}
    \label{prob:edma_token_power}
    \begin{align}
        & \underset{\boldsymbol{\vartheta},\mathbf A_\beta^\chi,\mathbf V_{\beta}^{\chi}}{\min}\quad \bar P^{\chi}
        \label{prob:edma_token_power_objective}\\
        & \mathrm{s.t.}\quad \gamma_{k,\beta}^{\chi}\geq\gamma_k^{\min}, k\in[K],\ \beta\in[B_{\rm f}],
        \label{prob:edma_token_sinr}\\
        & \Delta_k^{\chi}\leq\Delta_k^{\max}, k\in[K],
        \label{prob:edma_token_distortion}\\
        & \alpha_{j,\ell,\beta}^{\chi}\in\{0,1\}, \ell\in[L],\ j\in[K],\ \beta\in[B_{\rm f}],
        \label{prob:edma_token_binary}\\
        & \sum_{j=1}^{K}\alpha_{j,\ell,\beta}^{\chi}\leq 1, \ell\in[L],\ \beta\in[B_{\rm f}],
        \label{prob:edma_token_assignment}\\
        & v_{j,\ell,\beta}^{\chi}=0 \ \text{if }\alpha_{j,\ell,\beta}^{\chi}=0,
        \label{prob:edma_token_support}
\end{align}
\end{subequations}
where $\gamma_k^{\min}$ is the minimum SINR required by user $k$, and $\Delta_k^{\max}$ is the maximum admissible normalized token distortion. Constraint \eqref{prob:edma_token_sinr} maintains the prescribed physical link quality in every transmission interval, while \eqref{prob:edma_token_distortion} limits the expected error of the token block recovered over the frame. Constraints \eqref{prob:edma_token_binary} and \eqref{prob:edma_token_assignment} enforce exclusive element assignment, and \eqref{prob:edma_token_support} prevents a stream from using an unassigned element. 
Both designs $\pi_\beta^{\mathrm{sep}}$ and $\pi_\beta^{\mathrm{joint}}$ obey~\eqref{prob:edma_token_sinr}--\eqref{prob:edma_token_support}. The separate policy selects the assignments and beamformers from the environmental information alone. The joint policy also observes the current task-token blocks.

Problem~\eqref{prob:edma_token_power} is mixed-integer and generally nonconvex.
The assignments and beamformers are coupled through the support constraints. The beamformers also affect the effective links and action tokens supplied to the neural functions parameterized by $\boldsymbol{\vartheta}$. The token-distortion constraint therefore couples assignment and beamforming decisions through nonlinear encoding and decoding. Frame-level token recovery further couples the transmission intervals.
To address these couplings, Section~\ref{sec:algorithm} presents offline training of $\boldsymbol{\vartheta}$ and subsequent online assignment and beamforming optimization with fixed $\boldsymbol{\vartheta}$ for both designs.


\section{Two-Stage Optimization for EDMA-Aware Token Transmission Framework}
\label{sec:algorithm}

This section develops a two-stage approach with offline training and online optimization for both designs. Offline, the action-token encoder, fusion functions, symbol mapper, and receiver decoders are jointly trained over sampled task information and feasible physical configurations. With the trained parameters $\boldsymbol{\vartheta}$ fixed, TD-AO updates the element assignments and digital beamformers online under the SINR and average-distortion constraints. The computational complexity is then analyzed.

\subsection{Offline End-to-End Training}
\label{subsec:offline_training}

During offline training, the trainable components of token construction, token-to-symbol mapping, transmission and token recovery are jointly optimized through the complete transmission and recovery process. The training objective is evaluated from the reconstructed task tokens under the end-to-end learning framework~\cite{Bourtsoulatze2019DeepJSCC}. The communication modules involved in training are described below.

\subsubsection{Token Construction}
The task-token encoder $\Phi_{\rm task}(\cdot)$ is taken from a pretrained autoencoder with Kullback--Leibler (KL) regularization. It produces the task-token block $\mathbf Z_k$ and remains frozen while the communication modules are trained.
The action-token encoder $\Phi_{\rm act}(\cdot)$ uses a multi-layer perceptron (MLP) with two ReLU hidden layers of width $m_{\rm act}$ and a linear output layer of dimension $d_q$. $\Phi_{\rm act}(\cdot)$ processes the real-valued assignment and effective-channel inputs defined in~\eqref{eq:action_token}.
The projection $\Phi_{\rm EDMA}(\cdot)$ is a trainable affine map from $\mathbb R^{d_q}$ to $\mathbb R^{d_k}$ that matches the action-token dimension to the task-token dimension.
To fuse action tokens with task tokens, the residual function $\Phi_{\rm residual}(\cdot)$ uses an MLP with two ReLU hidden layers of width $m_{\rm res}$ and a linear output layer of dimension $d_k$. 
$\Phi_{\rm residual}(\cdot)$ takes the concatenated task token and projected action token as input. Equation~\eqref{eq:conditional_token_fusion} constructs the EDMA-aware tokens under the separate policy. Under the joint policy, the updated action tokens in~\eqref{eq:joint_action_token} are used, as described in Section~\ref{subsec:token_construction}. Both designs use the same subsequent transmission and recovery functions.


\subsubsection{Token-to-Symbol Mapping and Transmission}
To effectively transmit tokens, the token-to-symbol mapping $F_\mu(\cdot)$ is designed to convert EDMA-aware tokens into symbols $\mathbf s_{k,\beta}^{\chi}$. The token-to-symbol mapper consists of an input projection, $N_{\rm tx}$ transformer encoder blocks, a cross-attention layer, and a symbol output head. Specifically, the input projection maps each EDMA-aware token to $d_{\rm model}$ dimensions. Each transformer block uses $H_{\rm att}$ attention heads and a feed-forward sublayer of width $d_{\rm ff}$. Positional embeddings identify the token order. The cross-attention network uses $T_\beta$ symbol queries to obtain $T_\beta$ output features, and a linear head produces two real values for each query to form the complex symbol sequence $\mathbf s_{k,\beta}^{\chi}$.

\subsubsection{Token Recovery}
At the receiver, $D_{\phi,k}(\cdot)$ in~\eqref{eq:task_token_recovery} consists of an input projection, a transformer, and a task-token reconstruction head. The decoder uses $\mathcal I_{{\rm rx},k}^{\chi}$ and the received signal blocks $\{\mathbf y_{k,\beta}^{\chi}\}_{\beta=1}^{B_{\rm f}}$ to reconstruct the task-token block.
Specifically, the real and imaginary components of each received symbol $\mathbf y_{k,\beta}^{\chi}$ are augmented with the effective channel coefficients and noise variance of interval $\beta$.
These symbols are projected to $d_{\rm model}$ dimensions and processed by a transformer with $H_{\rm att}$ attention heads and a feed-forward sublayer of width $d_{\rm ff}$. The reconstruction head contains $N_k$ learnable output queries that correspond to the task-token positions in $\mathbf Z_k$.
These queries extract $N_k$ features from the encoded observations through cross-attention within $D_{\phi,k}(\cdot)$. A linear layer maps each extracted feature from dimension $d_{\rm model}$ to $d_k$, and the outputs form the reconstructed task-token block $\widehat{\mathbf Z}_k^\chi\in\mathbb R^{N_k\times d_k}$.

Let $\Delta_{k,{\rm tr}}(\boldsymbol{\vartheta})$ denote the expected distortion in~\eqref{eq:task_token_distortion} for user $k$ under the training distribution. The training expectation includes sampled task-source inputs, propagation conditions, assignments and beamformers, and receiver noise.
For a fixed sampled configuration, \eqref{eq:average_transmit_power} is independent of the trainable token-processing parameters because the symbol sequences are normalized. We therefore train these parameters offline to minimize the sum of the users' expected reconstruction errors, which is given by
\begin{equation}
    \label{prob:offline_token_training}
    \underset{\boldsymbol{\vartheta}}{\min} \ \sum_{k=1}^{K} \Delta_{k,{\rm tr}}(\boldsymbol{\vartheta}).
\end{equation}

Furthermore, the training expectation is approximated using mini-batches of $N_{\rm tr}$ frames. Let $\widehat{\Delta}_{k,{\rm tr}}^{(r)}$ denote the mini-batch mean distortion of user $k$ at iteration $r$. Mini-batch stochastic gradient descent updates the network parameters as follows:
\begin{equation}
    \label{eq:offline_parameter_update}
    \boldsymbol{\vartheta}^{(r+1)}
    =
    \boldsymbol{\vartheta}^{(r)}
    -
    \epsilon_{\rm tr}^{(r)}
    \sum_{k=1}^{K}
    \nabla_{\boldsymbol{\vartheta}}
    \widehat{\Delta}_{k,{\rm tr}}^{(r)}
    (\boldsymbol{\vartheta}^{(r)}),
\end{equation}
where $\epsilon_{\rm tr}^{(r)}>0$ is the learning rate. The reconstruction loss is backpropagated through the decoder, fixed channel operation, token-to-symbol mapping, residual-fusion network, action-feature projection, and action-token encoder. During this backpropagation, only their trainable parameters are updated, while task information, sampled channel coefficients, assignments, and beamformers remain fixed. Algorithm~\ref{alg:offline_training} summarizes the joint training of these modules for task-token recovery.

\begin{algorithm}[t]
\caption{Offline End-to-End Training}
\label{alg:offline_training}
\begin{algorithmic}[1]
\REQUIRE Frozen task encoder $\Phi_{\rm task}(\cdot)$, training data and physical-condition distribution, learning rates $\{\epsilon_{\rm tr}^{(r)}\}$, iteration limit $R_{\rm tr}$, mini-batch size $N_{\rm tr}$, $K$, $L$, $B_{\rm f}$, and $\chi$.
\ENSURE Trained network parameters $\boldsymbol{\vartheta}_{\rm tr}$.

\STATE Initialize $\boldsymbol{\vartheta}^{(0)}$.
\FOR{$r=0,\ldots,R_{\rm tr}-1$}
    \STATE Sample a mini-batch of $N_{\rm tr}$ complete frames.
    \FORALL{frames in the mini-batch}
        \STATE Extract $\mathcal Z=(\mathbf Z_1,\ldots,\mathbf Z_K)$ using $\Phi_{\rm task}(\cdot)$.
        \STATE Generate $\{\mathbf h_{k,\beta}\}_{k,\beta}$ using~\eqref{eq:user_channel_vector} under the frame-constant propagation model and form $\mathcal I_{\rm tx}$.
        \FOR{$\beta=1,\ldots,B_{\rm f}$}
            \IF{$\chi=\mathrm{sep}$}
                \STATE Generate $(\mathbf A_\beta^\chi,\mathbf V_\beta^\chi)$ using $\pi_\beta^{\rm sep}(\mathcal I_{\rm tx})$.
            \ELSE
                \STATE Generate $(\mathbf A_\beta^\chi,\mathbf V_\beta^\chi)$ using $\pi_\beta^{\rm joint}(\mathcal I_{\rm tx},\mathcal Z)$.
            \ENDIF
            \STATE Retain $(\mathbf A_\beta^\chi,\mathbf V_\beta^\chi)$ satisfying~\eqref{prob:edma_token_sinr},~\eqref{prob:edma_token_binary},~\eqref{prob:edma_token_assignment}, and~\eqref{prob:edma_token_support}.
            \STATE Compute $\{g_{k\leftarrow j,\beta}^\chi\}_{k,j=1}^{K}$ using~\eqref{eq:effective_coefficient}.
            \FOR{$k=1,\ldots,K$}
                \STATE Compute $\mathbf q_{k,\beta}^\chi$ using~\eqref{eq:action_token}.
                \STATE Construct $\widetilde{\mathbf Z}_{k,\beta}^\chi$ using~\eqref{eq:conditional_token_fusion} and~\eqref{eq:edma_aware_token_block}.
                \STATE Generate $\mathbf s_{k,\beta}^\chi$ using~\eqref{eq:token_symbol_mapping}.
            \ENDFOR
            \STATE Form $\mathbf X_\beta^\chi$ using~\eqref{eq:transmitted_block}.
            \STATE Generate independent receiver-noise samples.
            \STATE Compute $\{\mathbf y_{k,\beta}^\chi\}_{k=1}^{K}$ using~\eqref{eq:received_block}.
        \ENDFOR
        \FOR{$k=1,\ldots,K$}
            \STATE Form $\mathcal I_{{\rm rx},k}^\chi$ from effective coefficients and noise.
            \STATE Recover $\widehat{\mathbf Z}_k^\chi$ from $\{\mathbf y_{k,\beta}^\chi\}_{\beta=1}^{B_{\rm f}}$ and $\mathcal I_{{\rm rx},k}^\chi$ using~\eqref{eq:task_token_recovery}.
            \STATE Evaluated the distortion in~\eqref{eq:task_token_distortion} between $\mathbf Z_k$ and $\widehat{\mathbf Z}_k^\chi$.
        \ENDFOR
    \ENDFOR
    \STATE Compute the mini-batch mean distortions $\{\widehat{\Delta}_{k,{\rm tr}}^{(r)}\}_{k=1}^{K}$.
    \STATE Backpropagate the loss in~\eqref{prob:offline_token_training} with the sampled configurations fixed.
    \STATE Update $\boldsymbol{\vartheta}^{(r+1)}$ using~\eqref{eq:offline_parameter_update}.
\ENDFOR
\RETURN $\boldsymbol{\vartheta}_{\rm tr}=\boldsymbol{\vartheta}^{(R_{\rm tr})}$.
\end{algorithmic}
\end{algorithm}

\subsection{Online EDMA Assignment and Beamforming Optimization}
\label{subsec:online_optimization}

With the trained parameters $\boldsymbol{\vartheta}_{\rm tr}$ fixed, TD-AO alternates between digital beamformer refinement and discrete assignment updates. For fixed assignments, the beamformers are refined through inner iterations. A beamformer refinement followed by an assignment update constitutes an outer iteration. Under the frame-constant propagation assumption in Section~\ref{subsec:channel_model}, the channel vectors $\mathbf h_{k,\beta}(\rho_\beta,\mathbf b_{k,\beta})$ remain fixed during the configuration search.

For each candidate configuration, i.e., assignment and beamformer $\{(\mathbf A_\beta^\chi,\mathbf V_\beta^\chi)\}_{\beta=1}^{B_{\rm f}}$, the transmitter recomputes the effective coefficients $g_{k\leftarrow j,\beta}^{\chi}(\rho_\beta,\mathbf b_{k,\beta})$ and the corresponding action tokens $\mathbf q^{\chi}_{k,\beta}$, generates the symbol blocks $\mathbf s^{\chi}_{k,\beta}$, and evaluates token recovery $\Delta_k^\chi$. Candidate evaluation uses the complete frame because the decoder jointly processes observations from all intervals.
The designs use different information for candidate evaluation. For the separate design, $\widehat{\Delta}_{k}^{\mathrm{sep}}$ is averaged over reference task-token blocks and receiver-noise realizations conditional on $\mathcal I_{\rm tx}$. For the joint design, $\widehat{\Delta}_{k}^{\mathrm{joint}}$ is averaged over receiver noise conditional on $\mathcal I_{\rm tx}$ and the current task-token blocks $\mathcal Z$. These blocks remain fixed throughout the search and are used to evaluate both beamformer and assignment updates.
Monte Carlo sample averages approximate the respective expectations~\cite{Kleywegt2002SAA}. Each sample-average evaluation uses $M_{\rm MC}^{\rm sep}$ paired realizations of reference task-token blocks and receiver noise for the separate design, and $M_{\rm MC}^{\rm joint}$ receiver-noise realizations with the current task-token blocks $\mathcal Z$ fixed for the joint design. Each realization covers a complete frame. The same samples are retained throughout each configuration search.
Let $\widehat{\bar P}^{\chi}$ denote the conditional sample-average transmit power. To incorporate the average-distortion requirements into configuration search based on the conditional Lagrangian~\cite{Boyd2004ConvexOptimization}, we use the partial Lagrangian cost as follows:
\begin{equation}
    \label{eq:sample_average_candidate_cost}
    \widehat{\mathcal J}^{\chi}
    \triangleq
    \widehat{\bar P}^{\chi}
    +
    \sum_{k=1}^{K}
    \nu_k^\chi
    \bigl(
    \widehat{\Delta}_k^\chi-\Delta_k^{\max}
    \bigr),
\end{equation}
where $\nu_k^\chi\geq0$ is the nonnegative multiplier associated with user $k$'s average-distortion constraint.

\subsubsection{Digital Beamformer Update}
The average token-distortion constraints are incorporated into the sample-average partial Lagrangian in~\eqref{eq:sample_average_candidate_cost}, while the SINR and beamformer-support constraints remain explicit. With $\boldsymbol{\vartheta}_{\rm tr}$ and assignments $\{\mathbf A_\beta^\chi\}_{\beta=1}^{B_{\rm f}}$ fixed, the resulting beamforming problem minimizes $\widehat{\mathcal J}^{\chi}$ over $\{\mathbf v_{j,\beta}^{\chi}\}_{j,\beta}$ subject to~\eqref{prob:edma_token_sinr} and~\eqref{prob:edma_token_support}. It remains nonconvex because the objective depends nonlinearly on the beamformers through token construction, symbol mapping, and recovery, and the SINR constraints couple the desired and interfering links. Therefore, we construct a convex approximation at each beamformer iteration.
Specifically, let $\mathbf v_{j,\beta}^{\chi,(i)}$ denote the beamformer at the $i$-th iteration. The corresponding desired-link coefficient is given by
\begin{equation}
    \label{eq:current_desired_coefficient}
    g_{k\leftarrow k,\beta}^{\chi,(i)}=\mathbf h_{k,\beta}^{\mathsf H}(\rho_\beta,\mathbf b_{k,\beta})\mathbf v_{k,\beta}^{\chi,(i)}.
\end{equation}

The desired-signal power then admits the following global lower bound.

\begin{equation}
    \label{eq:desired_power_lower_bound}
    \begin{aligned}
        &\|\mathbf h_{k,\beta}^{\mathsf H}(\rho_\beta,\mathbf b_{k,\beta})\mathbf v_{k,\beta}^{\chi}\|^{2}\\
        & \geq 2\operatorname{Re}\{(g_{k\leftarrow k,\beta}^{\chi,(i)})^{*}\mathbf h_{k,\beta}^{\mathsf H}(\rho_\beta,\mathbf b_{k,\beta})\mathbf v_{k,\beta}^{\chi}\} - \|g_{k\leftarrow k,\beta}^{\chi,(i)}\|^{2}.
    \end{aligned}
\end{equation}

This bound follows from the convexity of the squared magnitude of a linear function and is tight at $\mathbf v_{k,\beta}^{\chi}=\mathbf v_{k,\beta}^{\chi,(i)}$. Substituting this lower bound into the desired-signal power in \eqref{prob:edma_token_sinr} gives a convex inner approximation of the SINR constraint. To construct a proximal first-order approximation of~\eqref{eq:sample_average_candidate_cost} under the successive convex approximation framework~\cite{Scutari2017SCA}, we first define
\begin{equation}
    \label{eq:beamformer_cost_derivative}
    \mathbf d_{j,\beta}^{\chi,(i)} \triangleq \left. \frac{\partial\widehat{\mathcal J}^{\chi}}{\partial(\mathbf v_{j,\beta}^{\chi})^{*}} \right|_{\{\mathbf v_{j,\beta}^{\chi,(i)}\}_{j,\beta}}.
\end{equation}

The derivative in~\eqref{eq:beamformer_cost_derivative} is evaluated at the current beamformers for all users and intervals and computed by backpropagation through the fixed functions.
Using these derivatives, we replace $\widehat{\mathcal J}^{\chi}$ by its first-order expansion with a quadratic proximal term. The constant objective value at the current iteration is omitted because it does not affect the minimization. By combining this objective approximation with SINR approximations~\cite{Scutari2017SCA}, the convex update problem is expressed as follows:
\begin{subequations}
    \label{prob:beamformer_sca_update}
    \begin{align}
        & \underset{\{\mathbf v_{j,\beta}^{\chi}\}_{j,\beta}}{\min} \ \sum_{\beta=1}^{B_{\rm f}}\sum_{j=1}^{K}[2\operatorname{Re}\{(\mathbf d_{j,\beta}^{\chi,(i)})^{\mathsf H}(\mathbf v_{j,\beta}^{\chi}-\mathbf v_{j,\beta}^{\chi,(i)})\} \nonumber \\
        & \qquad + \frac{\|\mathbf v_{j,\beta}^{\chi}-\mathbf v_{j,\beta}^{\chi,(i)}\|_2^2}{2\tau_i}]
        \label{prob:beamformer_sca_objective}\\
        &\mathrm{s.t.}\quad \gamma_k^{\min}(\sum_{j\ne k}|\mathbf h_{k,\beta}^{\mathsf H}(\rho_\beta,\mathbf b_{k,\beta})\mathbf v_{j,\beta}^{\chi}|^2 + \sigma_k^2)
        \nonumber\\
        & - 2\operatorname{Re}\{(g_{k\leftarrow k,\beta}^{\chi,(i)})^* \mathbf h_{k,\beta}^{\mathsf H}(\rho_\beta,\mathbf b_{k,\beta}) \mathbf v_{k,\beta}^{\chi}\}
        \nonumber\\
        & + |g_{k\leftarrow k,\beta}^{\chi,(i)}|^2 \leq 0, k\in[K],\ \beta\in[B_{\rm f}],
        \label{prob:beamformer_sca_sinr}\\
        & \eqref{prob:edma_token_support},
    \end{align}
\end{subequations}
where $\tau_i>0$ controls the update step. For fixed assignments, problem~\eqref{prob:beamformer_sca_update} is a convex quadratically constrained quadratic program. Constraint~\eqref{prob:beamformer_sca_sinr} is sufficient for the original SINR constraint~\eqref{prob:edma_token_sinr} because it uses a lower bound on the desired-signal power. Moreover, if beamformers satisfy the original SINR constraints, they also satisfy the approximated constraints constructed at that point.

After solving~\eqref{prob:beamformer_sca_update}, we evaluate the updated beamformers through the complete transmission and recovery process to obtain the sample-average objective in~\eqref{eq:sample_average_candidate_cost}. Backtracking reduces $\tau_i$ and solves~\eqref{prob:beamformer_sca_update} again until the objective decreases.
The same Monte Carlo samples are used throughout this comparison. The beamformer updates terminate when the relative objective change falls below a prescribed tolerance or the iteration limit is reached.

\subsubsection{EDMA Assignment Update}
\begin{algorithm}[t]
\caption{Token-Distortion-Guided Alternating Optimization (TD-AO) Algorithm for Online Optimization}
\label{alg:td_ao}
\begin{algorithmic}[1]
\REQUIRE $\chi$, $\mathcal I_{\rm tx}$, $\mathcal Z$, fixed parameters $\boldsymbol{\vartheta}_{\rm tr}$, calibrated multipliers $\{\nu_k^\chi\}_{k=1}^{K}$, improvement threshold $\epsilon_{\rm A}$, and search limit $R_{\rm A}$.
\ENSURE Assignments $\{\mathbf A_\beta^\chi\}_{\beta=1}^{B_{\rm f}}$ and beamformers $\{\mathbf v_{j,\beta}^\chi\}_{j,\beta}$.
\STATE Fix task-tokens and noise samples for $\chi=\mathrm{sep}$. For $\chi=\mathrm{joint}$, fix $\mathcal Z$ and noise samples.
\STATE Initialize $\{\mathbf A_\beta^\chi\}_{\beta=1}^{B_{\rm f}}$ using $\eta_{j,\ell,\beta}$ in~\eqref{eq:environment_selectivity}.
\STATE Obtain $\{\mathbf v_{j,\beta}^\chi\}_{j,\beta}$ satisfying~\eqref{prob:edma_token_sinr} and~\eqref{prob:edma_token_support}.
\STATE Refine $\{\mathbf v_{j,\beta}^\chi\}_{j,\beta}$ using the derivatives in~\eqref{eq:beamformer_cost_derivative}, the update in~\eqref{prob:beamformer_sca_update}, and backtracking.
\STATE Evaluate $\widehat{\mathcal J}^{\chi}$ using~\eqref{eq:sample_average_candidate_cost}.
\FOR{$r=1,\ldots,R_{\rm A}$}
    \STATE Set $\widehat{\mathcal J}_{\rm best}^{\chi}=\widehat{\mathcal J}^{\chi}$ and save the current configuration as the best candidate.
    \FOR{each neighboring assignment $\{\mathbf A_{\beta,\rm cand}^{\chi}\}_{\beta=1}^{B_{\rm f}}$}
        \STATE Obtain candidate beamformers $\{\mathbf v_{j,\beta,\mathrm{cand}}^\chi\}_{j,\beta}$ satisfying~\eqref{prob:edma_token_sinr} and~\eqref{prob:edma_token_support}.
        \IF{a SINR-feasible initialization is found}
            \STATE Refine $\{\mathbf v_{j,\beta,\rm cand}^{\chi}\}_{j,\beta}$ using~\eqref{eq:beamformer_cost_derivative}, \eqref{prob:beamformer_sca_update}, and backtracking.
            \STATE Recompute $\mathbf q_{k,\beta}^\chi$ and $\widetilde{\mathbf Z}_{k,\beta}^\chi$ in~\eqref{eq:action_token} and~\eqref{eq:conditional_token_fusion}.
            \STATE Generate $\mathbf s_{k,\beta}^\chi=F_\mu(\widetilde{\mathbf Z}_{k,\beta}^\chi)$, form $\mathbf X_\beta^\chi$ and $\mathbf y_{k,\beta}^\chi$ using~\eqref{eq:transmitted_block} and~\eqref{eq:received_block}.
            \STATE Recover $\widehat{\mathbf Z}_k^\chi$ in~\eqref{eq:task_token_recovery} and evaluate $\widehat{\mathcal J}_{\rm cand}^{\chi}$ in~\eqref{eq:sample_average_candidate_cost}.
            \IF{$\widehat{\mathcal J}_{\rm cand}^{\chi}<\widehat{\mathcal J}_{\rm best}^{\chi}$}
                \STATE Set $\widehat{\mathcal J}_{\rm best}^{\chi}=\widehat{\mathcal J}_{\rm cand}^{\chi}$.
                \STATE Save the candidate assignments and beamformers as the best configuration.
            \ENDIF
        \ENDIF
    \ENDFOR
    \IF{$\widehat{\mathcal J}_{\rm best}^{\chi}\leq\widehat{\mathcal J}^{\chi}-\epsilon_{\rm A}$}
        \STATE Replace $\{\mathbf A_\beta^\chi\}_{\beta=1}^{B_{\rm f}}$ and $\{\mathbf v_{j,\beta}^\chi\}_{j,\beta}$ with the best configuration and set $\widehat{\mathcal J}^{\chi}=\widehat{\mathcal J}_{\rm best}^{\chi}$.
    \ELSE
        \STATE \textbf{break}
    \ENDIF
\ENDFOR

\STATE Return $\{\mathbf A_\beta^\chi\}_{\beta=1}^{B_{\rm f}}$ and $\{\mathbf v_{j,\beta}^\chi\}_{j,\beta}$.
\end{algorithmic}
\end{algorithm}

Having established the beamformer update for fixed assignments $\{\mathbf A_\beta^\chi\}_{\beta=1}^{B_{\rm f}}$, we next optimize the assignments. This procedure is a greedy neighborhood search that evaluates local neighbors of the current assignment rather than all possible assignment matrices.

An initial assignment is obtained using the selectivity scores $\eta_{j,\ell,\beta}$. At each outer iteration, neighboring assignments $\mathbf A_{\beta,\mathrm{cand}}^\chi$ are generated by changing one element's association or exchanging two assignment rows. For example, reassigning element 2 from stream 1 to stream 2 changes its row from $[1,0]$ to $[0,1]$. To deactivate element $\ell$, set $\alpha_{j,\ell,\beta,\mathrm{cand}}^\chi=0$ for all $j\in[K]$ and retain the assignments of the other elements. Each element has at most $K$ alternative states. Together with two-element exchanges, these operations generate at most $B_{\rm f}(LK+L(L-1)/2)$ neighbors in each outer iteration.

For each $\mathbf A_{\beta,\mathrm{cand}}^\chi$, an SINR-feasible initialization is obtained by solving a minimum-norm beamforming problem under the prescribed assignment supports~\cite{Wiesel2006Conic}. The beamformers are then refined using~\eqref{prob:beamformer_sca_update}. In particular, $v_{j,\ell,\beta,\mathrm{cand}}^\chi=0$ whenever $\alpha_{j,\ell,\beta,\mathrm{cand}}^\chi=0$. A candidate is discarded if a feasible initialization satisfying $\gamma_{k,\beta}^\chi\geq\gamma_k^{\min}$ for all users and intervals cannot be obtained. Although a candidate changes the assignment in one interval, beamformer refinement accounts for the complete frame because the decoder jointly processes received observations from all intervals.

For each candidate assignment and its refined beamformers, the complete-frame transmission and recovery process is reevaluated to obtain $\widehat{\bar P}^\chi$ and $\widehat{\Delta}_k^\chi$. The candidate cost $\widehat{\mathcal J}_{\mathrm{cand}}^\chi$ is then computed using~\eqref{eq:sample_average_candidate_cost}.


The candidate with the lowest cost replaces the current configuration if $\widehat{\mathcal J}_{\mathrm{cand}}^\chi\leq\widehat{\mathcal J}^\chi-\epsilon_{\rm A}$, where $\widehat{\mathcal J}^\chi$ is the current cost and $\epsilon_{\rm A}$ is the minimum required improvement. Otherwise, the current configuration is retained.
The search terminates when no evaluated neighbor provides the required improvement or the prescribed search budget is exhausted. The candidates are evaluated using the same Monte Carlo task-token and noise samples, with the trained parameters $\boldsymbol{\vartheta}_{\rm tr}$ held fixed throughout the search.
Moreover, the distortion multipliers $\{\nu_k^\chi\}_{k=1}^{K}$ are calibrated offline and are held fixed throughout each online search.
At the $i$-th calibration iteration, the configuration-search procedure is run with fixed $\{\nu_k^{\chi,(i)}\}_{k=1}^{K}$, where $\nu_k^{\chi,(0)}\geq 0$ is initialized for every user.
The resulting assignments $\{\mathbf A_\beta^\chi\}_{\beta=1}^{B_{\rm f}}$ and beamformers $\{\mathbf v_{j,\beta}^\chi\}_{j,\beta}$ are used to evaluate token recovery $\Delta_k^\chi$ over the calibration realizations. Denote the corresponding sample-average distortion by $\widehat{\Delta}_{k,\mathrm{cal}}^{\chi,(i)}$.
The multipliers are updated according to the following:
\begin{equation}
    \label{eq:distortion_multiplier_update}
    \nu_k^{\chi,(i+1)} =[\nu_k^{\chi,(i)} + \epsilon_{\rm D}^{(i)}(\widehat{\Delta}_{k,\mathrm{cal}}^{\chi,(i)} - \Delta_k^{\max})]_{+},
\end{equation}
where $\epsilon_{\rm D}^{(i)}>0$ is the calibration step size and $[x]_{+}\triangleq\max\{x,0\}$. If $\widehat{\Delta}_{k,\mathrm{cal}}^{\chi,(i)}>\Delta_k^{\max}$, the multiplier $\nu_k^{\chi,(i+1)}$ increases. Thus, subsequent configuration searches place greater emphasis on user $k$'s reconstruction error. The configuration-search procedure is rerun with the updated multipliers.
Algorithm~\ref{alg:td_ao} summarizes the online TD-AO procedure.

\subsection{Computational Complexity Analysis}
\label{subsec:computational_complexity}

Offline training involves $R_{\rm tr}N_{\rm tr}$ complete-frame forward and backward evaluations, where $R_{\rm tr}$ and $N_{\rm tr}$ denote the number of training updates and the minibatch size, respectively. Both network training and distortion-multiplier calibration are completed before deployment.

For online optimization, each interval contains at most $L$ supported complex beamforming coefficients and $K$ SINR constraints. After gradient evaluation, the convex beamformer subproblem separates across intervals. Under a dense Newton implementation with at most $R_{\rm N}$ Newton steps per solve, its computational cost is $\mathcal O(R_{\rm N}B_{\rm f}(L^3+KL^2))$. The SINR-feasible initialization has the same order under this implementation. Let $R_{\rm BF}$ denote the maximum number of beamformer iterations and $R_{\rm ls}$ the maximum number of trial solves per iteration, including the initial trial. Single-element assignment changes and two-element exchanges generate at most $N_{\rm cand}\leq B_{\rm f}(LK+L(L-1)/2)$ neighbors per round.
Hence, with at most $R_{\rm A}$ search rounds, the computational cost of feasible initialization and convex subproblem solves in Algorithm~\ref{alg:td_ao} is bounded by $\mathcal O((1+R_{\rm A}N_{\rm cand})(1+R_{\rm BF}R_{\rm ls})R_{\rm N}B_{\rm f}(L^3+KL^2))$. This bound excludes the complete-frame neural forward evaluations used for candidate assessment and backtracking, and the backward evaluations used for gradient computation.


\section{Numerical Results}
\label{sec:numerical_results}


\begin{table}[t]
\centering
\caption{Default Simulation Parameters}
\label{tab:simulation_parameters}
\renewcommand{\arraystretch}{1.08}
\setlength{\tabcolsep}{3.2pt}
\begin{tabular}{p{0.48\columnwidth}
                p{0.18\columnwidth}
                p{0.25\columnwidth}}
\hline
\centering\textbf{Parameter} & \centering\textbf{Symbol} &
\centering\arraybackslash\textbf{Value}
\\
\hline
Number of transmit elements & \(L\) & \(16\)
\\
Number of element clusters & \(Q\) & \(4\)
\\
Elements per cluster & \(L/Q\) & \(4\)
\\
Element-cluster centers & \(\{\mathbf c_q\}_{q=1}^{Q}\) & \(\{(\pm9,\pm9,3)\}\) m
\\
Inter-element spacing & \(d_{\rm e}\) & \(\lambda/2\)
\\
Number of users & \(K\) & \(4\)
\\
User distribution region & \(\mathcal A_{\rm u}\) & \([-8,8]^2\) m\(^2\)
\\
User height & \(h_{\rm u}\) & \(1.5\) m
\\
Minimum inter-user distance & \(d_{\min}\) & \(1\) m
\\
Carrier frequency & \(f_{\rm c}\) & \(28\) GHz
\\
System bandwidth & \(W\) & \(1\) MHz
\\
Number of obstacles & \(N_{\rm obs}\) & \(4\)
\\
Obstacle-center region & \(\mathcal A_{\rm obs}\) & \([-6,6]^2\) m\(^2\)
\\
Obstacle length & \(d_{\rm obs}^{\rm len}\) & \(\mathcal U[2,4]\) m
\\
Obstacle thickness & \(d_{\rm obs}^{\rm th}\) & \(0.2\) m
\\
Obstacle height & \(h_{\rm obs}\) & \(2.5\) m
\\
Obstacle orientation & \(\theta_{\rm obs}\) & \(\mathcal U[0,\pi)\)
\\
Noise spectral density & \(N_0\) & \(-174\) dBm/Hz
\\
Receiver noise figure & \(F_{\rm N}\) & \(7\) dB
\\
Receiver noise power & \(\sigma_k^2\) & \(-107\) dBm
\\
Minimum SINR & \(\gamma_k^{\min}\) & \(0\) dB
\\
Number of intervals & \(B_{\rm f}\) & \(4\)
\\
Channel uses per interval & \(T_\beta\) & \(128\)
\\
Action-encoder hidden width & $m_{\rm act}$ & $128$\\
Residual-MLP hidden width & $m_{\rm res}$ & $128$\\
Transformer feature dimension & $d_{\rm model}$ & $128$\\
Feed-forward hidden width & $d_{\rm ff}$ & $512$\\
Number of attention heads & $H_{\rm att}$ & $4$\\
Number of mapper encoder blocks & $N_{\rm tx}$ & $2$\\
Separate-design evaluation samples & $M_{\rm MC}^{\rm sep}$ & $64$\\
Joint-design evaluation samples & $M_{\rm MC}^{\rm joint}$ & $64$\\
\hline
\end{tabular}
\end{table}

\subsection{Simulation Setup}
\label{subsec:simulation_setup}
We consider a multiuser downlink system with distributed transmit elements. In the implementation shown in Fig.~\ref{system_model}, each element has a dedicated RF chain that feeds a straight dielectric waveguide, and one antenna is selected for transmission.
Each radiating antenna is modeled as an isotropic radiator under the LoS channel model in~\eqref{eq:element_user_channel}. All elements are independently driven and can be assigned to different user streams~\cite{Ding2025EDMA,Ding2026AINativeEDMA}, according to the EDMA-aware action defined in Section~\ref{subsec:edma_aware_action}. For each physical realization, user locations are independently generated subject to a minimum separation constraint. Obstacles are then generated independently to determine the binary visibility indicator of each element-to-user link.
We use MS COCO 2017 with a seed-42 split of 113,287/5,000 training/validation images and all 5,000 images for testing. RGB images are resized along the shorter side to 256 pixels, center-cropped to $256\times256$, and normalized to $[-1,1]$. The frozen KL encoder produces unscaled posterior-mean latents of size $4\times 32 \times 32$. Flattening nonoverlapping $4\times4$ spatial patches across all channels in raster order gives $N_k=64$ tokens of dimension $d_k=64$.
The physical, channel, frame and network parameters are summarized in Table~\ref{tab:simulation_parameters}.

\begin{figure}[!t]
    \centering
    \includegraphics[width=0.4\textwidth]{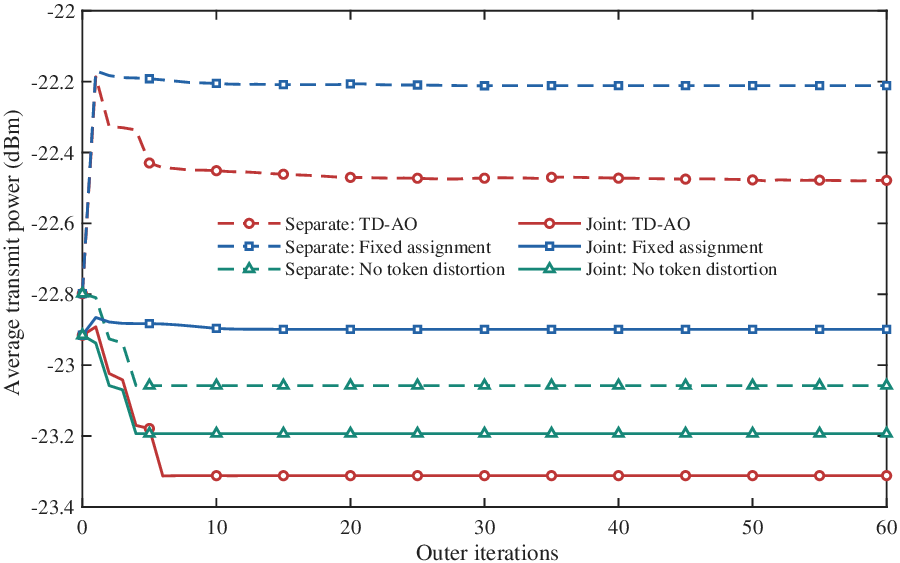}
    \caption{Average transmit power versus outer iteration for the separate and joint TD-AO designs and their reduced variants.}
    \label{fig:convergence}
\end{figure}

\begin{figure*}[t]
    \centering
    \subfloat[\label{fig:task_action_matching}]{%
        \includegraphics[width=0.31\textwidth]{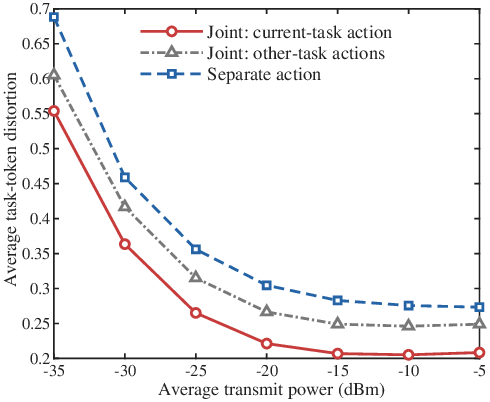}%
    }
    \hfill
    \subfloat[\label{fig:action_descriptor}]{%
        \includegraphics[width=0.32\textwidth]{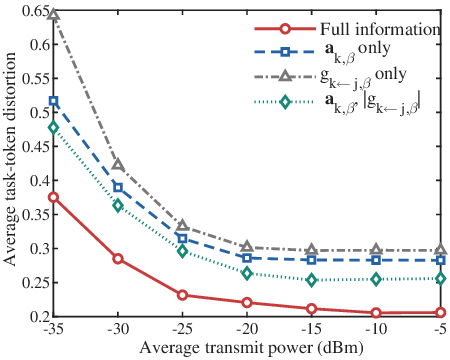}%
    }
    \hfill
    \subfloat[\label{fig:token_fusion}]{%
        \includegraphics[width=0.32\textwidth]{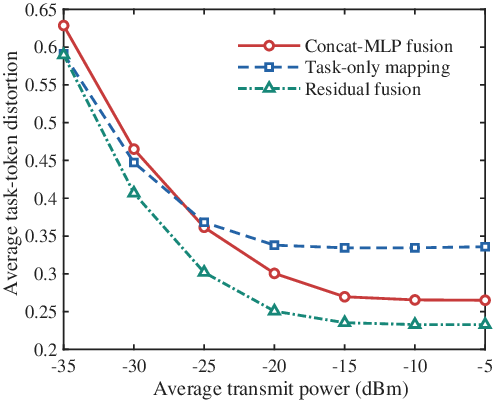}%
    }
    \caption{Performance comparison of average task-token distortion versus average transmit power for different components: (a) selection of the physical action, (b) information supplied to the action-token encoder, and (c) fusion of task and action tokens.}
    \label{fig:design_validation}
\end{figure*}

\subsection{Comparison Schemes}
\label{subsec:comparison_schemes}


We compare the proposed TD-AO algorithm under joint and separate designs with conventional schemes as follows:
\begin{itemize}
    \item \emph{Conv ZF:} The conventional zero-forcing scheme, denoted by Conv ZF, first constructs a fixed EDMA assignment using the channel-based element-selectivity metric in Section~II-B. Subject to the resulting assignment support, the digital beamforming directions are designed to suppress inter-user interference according to the ZF criterion~\cite{Spencer2004ZF}. The stream powers are subsequently adjusted to satisfy the prescribed SINR constraints with minimum total transmit power. Neither the element assignment nor the beamforming directions are optimized using task-token distortion.
    \item \emph{Conv MRT:} The maximum-ratio-transmission scheme, denoted by Conv MRT, uses the same fixed channel-based EDMA assignment as Conv ZF. For each stream, the beamforming direction is aligned with the intended user's channel over the elements assigned to that stream~\cite{Lo1999MRT}. The stream powers are then determined under the same SINR and total-power definitions used for other schemes.
\end{itemize}

For Conv ZF and Conv MRT, each transmit element is initially assigned according to the environment-selectivity metric $\eta_{j,\ell,\beta}$. In particular, an eligible element is assigned to the stream for which its desired-to-leakage selectivity is largest, subject to the exclusive-assignment constraint. The resulting assignments remain fixed during subsequent beamformer and power optimization. Consequently, these two schemes use instantaneous channel information but do not exploit the recovered task-token distortion to modify the EDMA divisions.

All schemes use the same physical channel realizations, task-token blocks, receiver-noise samples, frame structure, and total transmit-power definition. Power and distortion are compared only for channel realizations in which all schemes satisfy the SINR and assignment-support constraints. SINR-feasibility rates are evaluated over the full test set.
Unless otherwise stated, they also employ the same trained token-to-symbol mapper and token-recovery network. The joint and separate designs differ only in whether the current task-token blocks are available for physical-configuration selection at the transmitter.

\subsection{Convergence Behavior and Design Validation}
\label{subsec:convergence_validation}

This subsection first examines the convergence of the proposed online algorithm. It then evaluates the effects of task--action matching, action-descriptor construction, and token fusion to assess the contribution of each component.


Fig.~\ref{fig:convergence} shows the average transmit power over the outer iterations for TD-AO. The joint TD-AO design converges within approximately ten outer iterations. The separate design exhibits its largest changes during the first few iterations, followed by smaller adjustments. The initial power increase observed for some variants indicates that the power sequence is not necessarily monotonic during the search.
In particular, the joint TD-AO design reaches approximately $-23.31$~dBm, compared with $-22.48$~dBm for the separate design. This corresponds to a $17.4\%$ reduction in transmit power.
The online updates account for task-token distortion as well as transmit power, so an update that improves the distortion performance can require additional power.
The fixed-assignment variants converge to higher power levels than the corresponding full TD-AO designs, which indicates that beamformer refinement alone cannot fully compensate for a restricted assignment. Furthermore, removing token-distortion-guided refinement also increases the transmit power of the joint design.
Fig.~\ref{fig:design_validation} evaluates the contributions of the framework components. Specifically, Fig.~\ref{fig:design_validation}(a) compares EDMA-aware actions selected using different task information. The current-task action scheme uses the joint action selected for the current task-token blocks and achieves the lowest distortion throughout the considered power range. For the same physical channel, the other-task action scheme applies joint actions selected for other task realizations. Its advantage over the separate design suggests that statistical similarities across task-token blocks can make actions selected for other tasks beneficial relative to actions based solely on channel information. The further distortion reduction achieved by the current-task action demonstrates the additional benefit of matching the physical configuration to the specific token content.


Fig.~\ref{fig:design_validation}(b) compares the full action descriptor with three reduced inputs: assignment information only $\mathbf{A}_\beta$, complex effective coupling only $g_{k\leftarrow j,\beta}$, and assignment information $\mathbf{A}_\beta$ combined with coupling magnitudes $|g_{k\leftarrow j,\beta}|$.
The combination of assignment information $\mathbf{A}_\beta$ and coupling magnitudes $|g_{k\leftarrow j,\beta}|$ yields lower distortion than either assignment information $\mathbf{A}_\beta$ or complex effective coupling $g_{k\leftarrow j,\beta}$ alone. This result indicates that resource assignments and the resulting propagation conditions provide complementary information for token encoding.
For the same assignment input, retaining the phases of the effective channels further reduces token-recovery distortion.

Fig.~\ref{fig:design_validation}(c) compares residual fusion, concatenation followed by an MLP, and task-only mapping. Residual fusion yields the lowest distortion across the tested power range. The persistent gap at high transmit powers further supports the benefit of preserving task information when token construction is conditioned on the physical configuration.

\subsection{Communication and Recovery Performance}
\label{subsec:power_tradeoffs}

\begin{figure*}[t]
    \centering
    \subfloat[\label{fig:distortion_power}]{
        \includegraphics[width=0.31\textwidth]{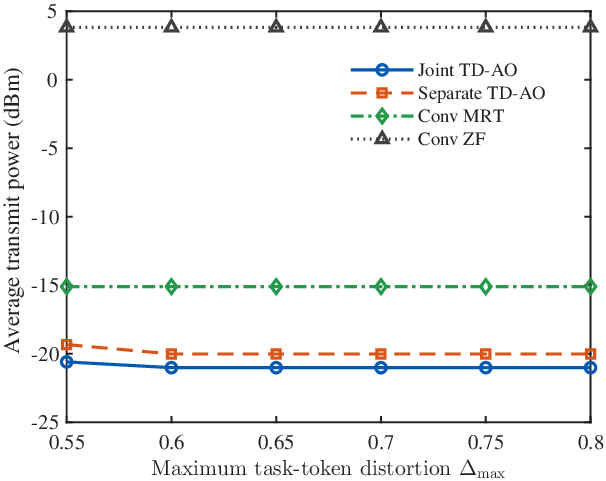}
    }
    \hfill
    \subfloat[\label{fig:sinr_power}]{
        \includegraphics[width=0.305\textwidth]{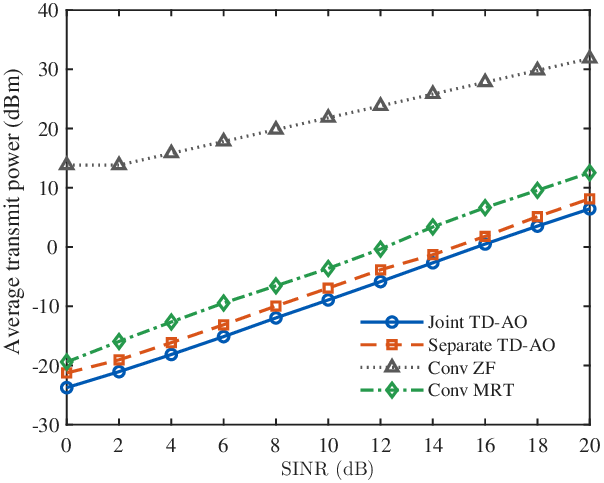}
    }
    \hfill
    \subfloat[\label{fig:blockage_robustness}]{
        \includegraphics[width=0.31\textwidth]{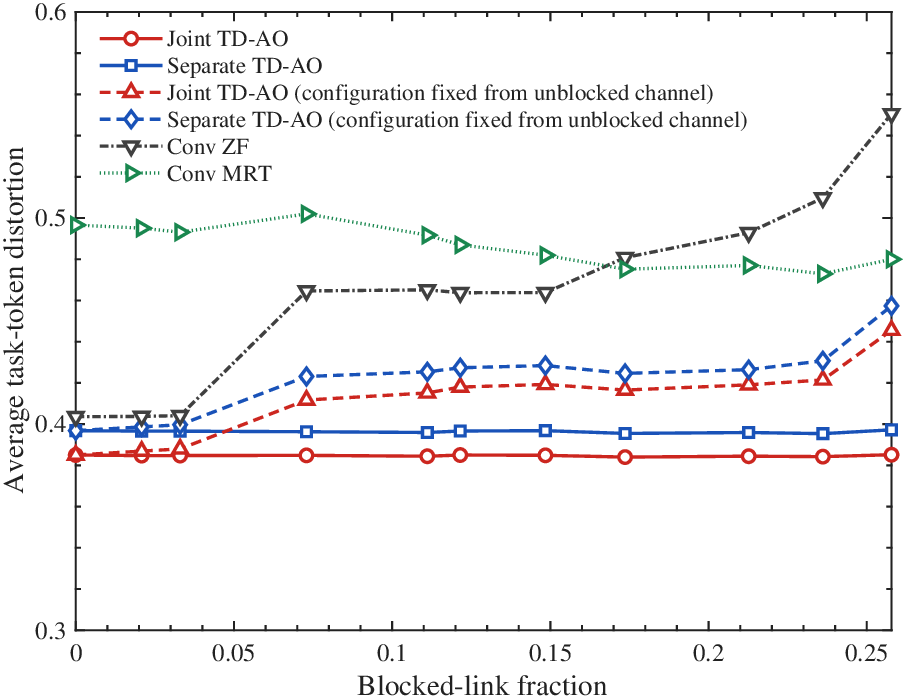}
    }
    \caption{Performance under different recovery requirements and channel conditions: (a) average transmit power versus maximum task-token distortion, with $\Delta_k^{\max}=\Delta_{\max}$ for all users; (b) average transmit power versus minimum SINR requirement; and (c) average task-token distortion versus blocked-link fraction under a transmit-power budget of $10$~dBm.}
    \label{fig:performance_comparison}
\end{figure*}

\begin{figure*}[t]
    \centering
    \subfloat[\label{fig:channeluse_power}]{%
        \includegraphics[width=0.32\textwidth]{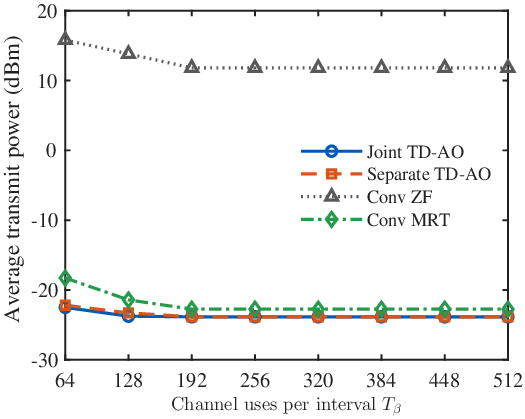}%
    }
    \hfill
    \subfloat[\label{fig:user_power}]{%
        \includegraphics[width=0.32\textwidth]{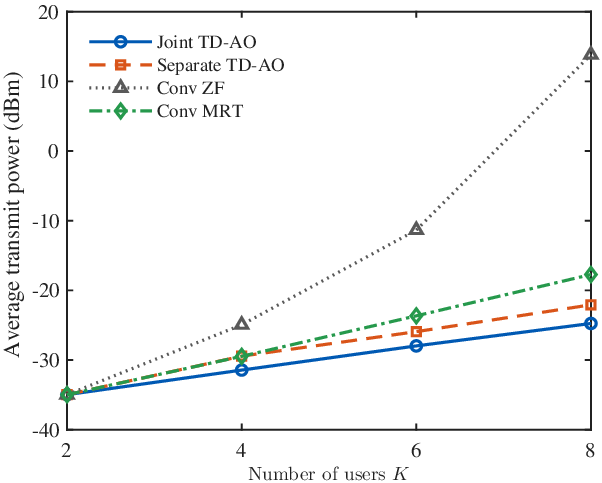}%
    }
    \hfill
    \subfloat[\label{fig:element_power}]{%
        \includegraphics[width=0.32\textwidth]{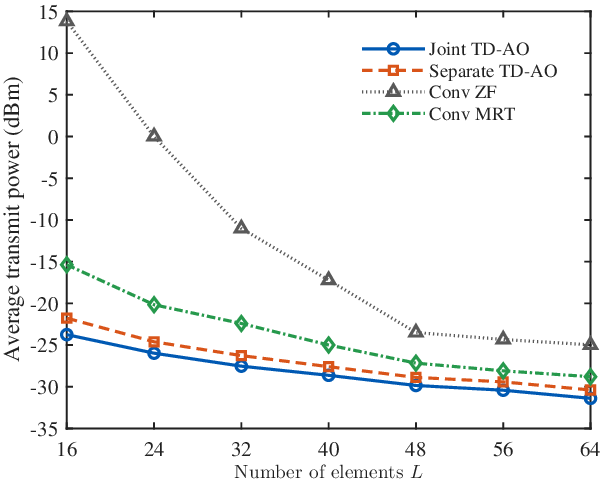}%
    }
    \caption{Average transmit power versus (a) the number of channel uses per interval, (b) the number of users, and (c) the number of transmit elements.}
    \label{fig:resources_user_load}
\end{figure*}

Fig.~\ref{fig:performance_comparison} evaluates the effects of the token-distortion requirement, SINR requirement, and link blockage on transmission performance.
Fig.~\ref{fig:performance_comparison}(a) illustrates the effect of the token-distortion requirement on transmit power. Increasing $\Delta_{\max}$ from $0.55$ to $0.60$ reduces the power required by both TD-AO designs, since a relaxed recovery requirement permits more power-efficient assignments and beamformers. Joint TD-AO consistently requires less power than separate TD-AO, which supports matching the physical action to the current task-token realization. Further relaxation yields little additional reduction. Thus, the benefit diminishes under the remaining transmission constraints. In contrast, Conv MRT and Conv ZF exhibit nearly constant power because their physical configurations are selected without using the token-distortion threshold. These results show that incorporating recovery requirements into physical-action selection enables TD-AO to exploit the permitted distortion tolerance for power reduction.


Fig.~\ref{fig:performance_comparison}(b) compares the required transmit power under different minimum SINR requirements. The required power increases for all schemes as the SINR threshold rises because stronger desired signals are required relative to interference and noise.
Conv MRT strengthens the desired links but does not explicitly eliminate inter-user interference. Conv ZF suppresses interference through channel nulling, which restricts the available beamforming directions and can reduce the desired-link gain. Instead, TD-AO adjusts the assignments and beamformers to balance desired-signal enhancement and interference suppression.
The persistent advantage of the joint design over the separate design shows that current-task information remains beneficial even under stringent SINR requirements.


Fig.~\ref{fig:performance_comparison}(c) compares the average task-token distortion at different blocked-link fractions under a transmit-power budget of $10$~dBm. The blocked-link fraction is defined as the number of blocked element--user links divided by $LK$, averaged over the test channel realizations. Importantly, blocking an individual link does not necessarily disconnect a user because its stream can still be transmitted through other elements. Therefore, the effect of blockage depends on which links are removed and whether the physical configuration is updated to exploit the remaining links.
Both blockage-aware TD-AO designs maintain nearly constant distortion, whereas their counterparts retaining configurations selected for unblocked channels exhibit increasing distortion. Updating the assignments and beamformers allows transmission to exploit the remaining visible links, while the corresponding action tokens inform symbol generation of the updated configuration. The joint design also preserves its lower distortion across the considered range. These results demonstrate that configuration adaptation improves robustness to blockage. Task-dependent selection provides an additional recovery gain.

Fig.~\ref{fig:resources_user_load} shows how the number of channel uses, users, and transmit elements affects the average transmit power.
Fig.~\ref{fig:resources_user_load}(a) shows the average transmit power versus the number of data channel uses in each interval. Increasing $T_\beta$ from $64$ to approximately $192$ reduces the average transmit power required by both TD-AO designs, whereas further increases yield little additional reduction. Longer symbol sequences provide greater flexibility for conveying the same task-token blocks under the recovery requirement.
The joint and separate curves also become nearly indistinguishable as $T_\beta$ increases. This suggests that additional channel uses reduce the average-power penalty associated with configuration selection without current-task information.

Fig.~\ref{fig:resources_user_load}(b) illustrates the effect of the number of users on the average transmit power. All schemes require more power as $K$ increases because more streams compete for the available elements and generate additional interference. The schemes achieve similar power levels at $K=2$, while their differences become more pronounced as the user load grows. With a fixed number of transmit elements, a higher user load reduces the average number of elements available to each stream and imposes more interference-nulling conditions. These effects contribute to the steep power increase observed for Conv ZF. Conv MRT also requires more power than TD-AO, as its channel-aligned beamforming directions do not explicitly account for inter-user interference or token-recovery distortion. In particular, the widening gap between joint and separate TD-AO designs shows that current-task information becomes more valuable when element allocation is more restrictive. Therefore, EDMA-aware configuration improves power efficiency under increased resource competition.


Fig.~\ref{fig:resources_user_load}(c) shows the effect of the number of elements on the average transmit power. All schemes require less power as $L$ increases. Additional elements offer greater flexibility in forming user-specific divisions and designing digital beamformers. They can also strengthen desired links and provide more options for interference suppression. Conv ZF benefits particularly strongly. This is consistent with the ability of additional spatial resources to alleviate the cost of interference nulling. The power gap between joint and separate TD-AO narrows as $L$ increases because configuration selection also becomes more effective with additional elements. This comparison shows that the benefit of using current-task information for configuration selection is most pronounced when users compete for limited spatial resources.

\section{Conclusion}
\label{sec:conclusion}
In this paper, we investigated the interaction between token construction and EDMA configuration. Our results indicate that the value of task-dependent configuration depends strongly on resource availability. Matching element assignments and beamformers to the current task provides greater power savings when more users compete for limited transmit elements, whereas additional elements and channel uses narrow the gap between the joint and separate designs. This dependence suggests that task information is particularly valuable for allocating scarce transmission resources. The component comparisons further show that element assignments and complex effective channels provide complementary information for token construction. An effective action representation should therefore describe both the selected resources and the resulting desired and interfering links, and retain channel phase information. The blockage experiments also highlight the importance of configuration adaptation: updates to assignments and beamformers help maintain recovery quality through the remaining visible links. Together, these findings support a coordinated design in which current task content guides physical configuration and the resulting configuration guides token construction.

Numerical results have shown that matching EDMA assignments and digital beamformers to the task-token realization provides additional power savings over the separate design. This benefit has become more pronounced as more users compete for limited transmit elements, while additional channel uses and transmit elements have reduced the gap between the two designs. Furthermore, action tokens that encode both the element assignments and the complex effective channels yield lower token-recovery distortion than tokens that encode either input alone.

\bibliographystyle{IEEEtran}
\def\baselinestretch{1}
\bibliography{TokenCom}

@article{Ding2025EDMA,
  author  = {Ding, Zhiguo and Schober, Robert and Poor, H. Vincent},
  title   = {Environment Division Multiple Access ({EDMA}): A Feasibility Study via Pinching Antennas},
  journal = {IEEE Trans. Wireless Commun.},
  volume  = {25},
  pages   = {15675--15691},
  year    = {2026},
  doi     = {10.1109/TWC.2026.3683028}
}

@article{Ding2026AINativeEDMA,
  author  = {Zhiguo Ding},
  title   = {Environment-Division Multiple Access: An Enabler for {AI}-Native Multiple Access},
  journal = {Engineering},
  year    = {2026},
  month   = {early access,},
  doi     = {10.1016/j.eng.2026.08.017}
}

@article{Shao2022TaskOriented,
  author  = {Jiawei Shao and Yuyi Mao and Jun Zhang},
  title   = {Learning Task-Oriented Communication for
             Edge Inference: An Information Bottleneck
             Approach},
  journal = {IEEE J. Sel. Areas Commun.},
  volume  = {40},
  number  = {1},
  pages   = {197--211},
  month   = jan,
  year    = {2022},
  doi     = {10.1109/JSAC.2021.3126087}
}

@article{Qiao2025TokenCommunications,
  author  = {Li Qiao and Mahdi Boloursaz Mashhadi
             and Zhen Gao and Rahim Tafazolli
             and Mehdi Bennis and Dusit Niyato},
  title   = {Token Communications: A Large Model-Driven
             Framework for Cross-Modal Context-Aware
             Semantic Communications},
  journal = {IEEE Wireless Commun.},
  volume  = {32},
  number  = {5},
  pages   = {80--88},
  month   = oct,
  year    = {2025},
  doi     = {10.1109/MWC.001.2500084}
}

@article{Zeng2021CKM,
  author  = {Yong Zeng and Xiaoli Xu},
  title   = {Toward Environment-Aware {6G} Communications
             via Channel Knowledge Map},
  journal = {IEEE Wireless Commun.},
  volume  = {28},
  number  = {3},
  pages   = {84--91},
  month   = jun,
  year    = {2021},
  doi     = {10.1109/MWC.001.2000327}
}

@article{Dai2022NTSCC,
  author  = {Jincheng Dai and Sixian Wang and Kailin Tan
             and Zhongwei Si and Xiaoqi Qin and Kai Niu
             and Ping Zhang},
  title   = {Nonlinear Transform Source-Channel Coding
             for Semantic Communications},
  journal = {IEEE J. Sel. Areas Commun.},
  volume  = {40},
  number  = {8},
  pages   = {2300--2316},
  month   = aug,
  year    = {2022},
  doi     = {10.1109/JSAC.2022.3180802}
}

@article{Ying2026TokenCommunications,
  author  = {Jingkai Ying and Zhijin Qin and Yulong Feng
             and Liejun Wang and Xiaoming Tao},
  title   = {Joint Semantic-Channel Coding and Modulation
             for Token Communications},
  journal = {IEEE Trans. Wireless Commun.},
  volume  = {25},
  pages   = {8179--8193},
  year    = {2026},
  doi     = {10.1109/TWC.2025.3636174}
}

@article{Zhu2024MAModel,
  author  = {Lipeng Zhu and Wenyan Ma and Rui Zhang},
  title   = {Modeling and Performance Analysis for Movable Antenna Enabled Wireless Communications},
  journal = {IEEE Trans. Wireless Commun.},
  volume  = {23},
  number  = {6},
  pages   = {6234--6250},
  month   = jun,
  year    = {2024},
  doi     = {10.1109/TWC.2023.3330887}
}

@article{Erak2026AdaptiveTokenMerging,
  author  = {Omar Erak and Omar Alhussein and Hatem Abou-Zeid and Mehdi Bennis and Sami Muhaidat},
  title   = {Adaptive Token Merging for Efficient Transformer Semantic Communication at the Edge},
  journal = {IEEE Open J. Commun. Soc.},
  volume  = {7},
  pages   = {4112--4128},
  month   = mar,
  year    = {2026},
  doi     = {10.1109/OJCOMS.2026.3676928}
}

@article{Shin2026ContextAwareToken,
  author  = {Junyong Shin and Joohyuk Park and Jihong Park and Jinho Choi and Yo-Seb Jeon},
  title   = {Context-Aware Iterative Token Detection and Masked Transmission for Wireless Token Communication},
  month   = Jan,
  journal = {arXiv preprint arXiv:2601.17770},
  year    = {2026}
}

@article{Zhang2026AdaTokenCom,
  author  = {Zijun Zhang and Li Qiao and Mahdi Boloursaz Mashhadi and Zhen Gao and Mehdi Bennis and Kaibin Huang},
  title   = {{Ada-TokenCom}: Rate-Adaptive Token Communications via Large-Model-Driven Token Compression and Generation},
  month   = Aug,
  journal = {arXiv preprint arXiv:2608.28086},
  year    = {2026}
}

@article{Zhao2026SiteSpecificMIMO,
  author  = {Cheng-Jie Zhao and Zhaolin Wang and Zongyao Zhao and Yuanwei Liu},
  title   = {Site-Specific Learning for Low-Overhead Multi-User {MIMO} Beamforming},
  month   = Jul,
  journal = {arXiv preprint arXiv:2607.10746},
  year    = {2026}
}

@article{Devoto2026AdaptiveToken,
  author  = {Alessio Devoto and Jary Pomponi
             and Mattia Merluzzi and Paolo Di Lorenzo
             and Simone Scardapane},
  title   = {Adaptive Semantic Token Communication
             for Transformer-Based Edge Inference},
  journal = {IEEE Trans. Mach. Learn. Commun. Netw.},
  volume  = {4},
  pages   = {422--437},
  year    = {2026},
  doi     = {10.1109/TMLCN.2026.3659819}
}

@article{Wang2026GenerativeSSBF,
  author  = {Zhaolin Wang and Zihao Zhou
             and Cheng-Jie Zhao and Yuanwei Liu},
  title   = {Generative Site-Specific Beamforming
             for Next-Generation Spatial Intelligence},
  journal = {IEEE Commun. Mag.},
  note    = {Early access},
  year    = {2026},
  doi     = {10.1109/MCOM.001.2600001}
}

@article{Sadek2007Leakage,
  author  = {Mirette Sadek and Alireza Tarighat and Ali H. Sayed},
  title   = {A Leakage-Based Precoding Scheme for Downlink Multi-User {MIMO} Channels},
  journal = {IEEE Trans. Wireless Commun.},
  volume  = {6},
  number  = {5},
  pages   = {1711--1721},
  month   = may,
  year    = {2007}
}

@article{Sadek2007Antenna,
  author  = {Mirette Sadek and Alireza Tarighat and Ali H. Sayed},
  title   = {Active Antenna Selection in Multiuser {MIMO} Communications},
  journal = {IEEE Trans. Signal Process.},
  volume  = {55},
  number  = {4},
  pages   = {1498--1510},
  month   = apr,
  year    = {2007},
  doi     = {10.1109/TSP.2006.888893}
}

@article{HajHmida2021Pilot,
  author  = {Wafa Haj Hmida and Vahid Meghdadi and Ammar Bouallegue and Jean Pierre Cances},
  title   = {Graph-Colouring Based Pilot Assignment to Mitigate Downlink Pilot Contamination for Cell-Free Massive {MIMO} Systems},
  journal = {IET Communications},
  volume  = {15},
  number  = {16},
  pages   = {2125--2138},
  year    = {2021},
  doi     = {10.1049/cmu2.12245}
}

@article{Bourtsoulatze2019DeepJSCC,
  author  = {Eirina Bourtsoulatze and David Burth Kurka and Deniz G{\"u}nd{\"u}z},
  title   = {Deep Joint Source-Channel Coding for Wireless Image Transmission},
  journal = {IEEE Trans. Cogn. Commun. Netw.},
  volume  = {5},
  number  = {3},
  pages   = {567--579},
  year    = {2019},
  doi     = {10.1109/TCCN.2019.2919300}
}

@article{Wiesel2006Conic,
  author  = {Ami Wiesel and Yonina C. Eldar and Shlomo Shamai},
  title   = {Linear Precoding via Conic Optimization for Fixed {MIMO} Receivers},
  journal = {IEEE Trans. Signal Process.},
  volume  = {54},
  number  = {1},
  pages   = {161--176},
  month   = jan,
  year    = {2006},
  doi     = {10.1109/TSP.2005.861073}
}

@article{Kleywegt2002SAA,
  author  = {Anton J. Kleywegt and Alexander Shapiro
             and Tito Homem-de-Mello},
  title   = {The Sample Average Approximation Method
             for Stochastic Discrete Optimization},
  journal = {SIAM Journal on Optimization},
  volume  = {12},
  number  = {2},
  pages   = {479--502},
  year    = {2002},
  doi     = {10.1137/S1052623499363220}
}

@article{Scutari2017SCA,
  author  = {Gesualdo Scutari and Francisco Facchinei
             and Lorenzo Lampariello},
  title   = {Parallel and Distributed Methods for
             Constrained Nonconvex Optimization--{P}art {I}:
             Theory},
  journal = {IEEE Trans. Signal Process.},
  volume  = {65},
  number  = {8},
  pages   = {1929--1944},
  month   = apr,
  year    = {2017},
  doi     = {10.1109/TSP.2016.2637317}
}

@book{Boyd2004ConvexOptimization,
  author    = {Stephen Boyd and Lieven Vandenberghe},
  title     = {Convex Optimization},
  publisher = {Cambridge University Press},
  address   = {Cambridge, U.K.},
  year      = {2004}
}

@article{Spencer2004ZF,
  author  = {Q. H. Spencer and A. L. Swindlehurst and M. Haardt},
  title   = {Zero-Forcing Methods for Downlink Spatial Multiplexing
             in Multiuser {MIMO} Channels},
  journal = {IEEE Trans. Signal Process.},
  volume  = {52},
  number  = {2},
  pages   = {461--471},
  month   = feb,
  year    = {2004},
  doi     = {10.1109/TSP.2003.821107}
}

@article{Lo1999MRT,
  author  = {T. K. Y. Lo},
  title   = {Maximum Ratio Transmission},
  journal = {IEEE Trans. Commun.},
  volume  = {47},
  number  = {10},
  pages   = {1458--1461},
  month   = oct,
  year    = {1999},
  doi     = {10.1109/26.795811}
}

@article{Wong2021FAS,
  author  = {Kai-Kit Wong and Arman Shojaeifard and Kin-Fai Tong and Yangyang Zhang},
  title   = {Fluid Antenna Systems},
  journal = {IEEE Trans. Wireless Commun.},
  year    = {2021},
  volume  = {20},
  number  = {3},
  pages   = {1950--1962},
  month   = mar,
  doi     = {10.1109/TWC.2020.3037595}
}

@article{Wang2025Modeling,
  author  = {Zhaolin Wang and Chongjun Ouyang and Xidong Mu and Yuanwei Liu and Zhiguo Ding},
  title   = {Modeling and Beamforming Optimization for Pinching-Antenna Systems},
  journal = {IEEE Trans. Commun.},
  year    = {2025},
  volume  = {73},
  number  = {12},
  pages   = {13904--13919},
  month   = dec,
  doi     = {10.1109/TCOMM.2025.3621049}
}

@article{Ouyang2026RateRegion,
  author  = {Chongjun Ouyang and Zhaolin Wang and Yuanwei Liu
             and Zhiguo Ding},
  title   = {Rate Region of {ISAC} for Pinching-Antenna Systems},
  journal = {IEEE Trans. Commun.},
  month   = {early access,},
  year    = {2026},
  doi     = {10.1109/TCOMM.2026.3668164}
}

\end{document}